\documentclass[a4paper,aps,prd,twocolumn,showpacs,longbibliography]{revtex4-1}
\pdfoutput=1

\usepackage{enumerate}
\usepackage{graphicx}
\usepackage[caption=false]{subfig}
\usepackage{hyperref}

\usepackage{bm}
\usepackage{amsmath}
\usepackage{amssymb}

\newcommand{\Ans}[1][n]{A_{#1}^{*}}

\newcommand{\Fstat}{^{0}}
\newcommand{\Ufit}{^{\textrm{fit}}}
\newcommand{\Zn}[1][n]{\mathbb{Z}^{#1}}
\newcommand{\calA}{\mathcal{A}}
\newcommand{\calF}{\mathcal{F}}
\newcommand{\calP}{\mathcal{P}}
\newcommand{\cohsmat}[1]{\cohs{\mat{#1}}}
\newcommand{\cohsvec}[1]{\cohs{\vec{#1}}}
\newcommand{\cohs}[1]{\coh{#1}\useg}
\newcommand{\coh}[1]{\widetilde{#1}}
\newcommand{\diff}{\Delta}

\newcommand{\mat}[1]{\boldsymbol{\mathrm{#1}}}
\newcommand{\mean}[1]{\langle #1 \rangle}
\newcommand{\ndot}[2][s]{#2^{(#1)}}
\newcommand{\phase}{^{\phi}}
\newcommand{\semimat}[1]{\semi{\mat{#1}}}
\newcommand{\semivec}[1]{\semi{\vec{#1}}}
\newcommand{\semi}[1]{\widehat{#1}}
\newcommand{\sigvec}[1]{\sig{\vec{#1}}}
\newcommand{\sig}[1]{#1^\mathrm{s}}
\newcommand{\stdv}[1]{\sigma_{#1}}
\newcommand{\sums}[1][N]{\sum_{\ell=1}^{#1}}
\newcommand{\ua}{_{a}}
\newcommand{\ub}{_{b}}
\newcommand{\umax}{_{\textrm{max}}}
\newcommand{\unoint}{_\textrm{ni}}
\newcommand{\useg}{_{\ell}}
\renewcommand{\vec}[1]{\boldsymbol{#1}}

\newcommand{\gitDate}{2016-12-02 15:06:10 +0100}
\newcommand{\gitID}{df432c9}
\newcommand{\gitStatus}{CLEAN}

\begin{document}

\title{Empirically extending the range of validity of parameter-space \\ metrics for all-sky searches for gravitational-wave pulsars}
\author{Karl Wette}
\email{karl.wette@aei.mpg.de}
\affiliation{Max Planck Institute for Gravitational Physics (Albert Einstein Institute) and Leibniz Universit\"at Hannover, Callinstra\ss{}e 38, 30167 Hannover, Germany}

\date{\gitDate, commit \gitID-\gitStatus}

\begin{abstract}
All-sky searches for gravitational-wave pulsars are generally limited in sensitivity by the finite availability of computing resources.
Semicoherent searches are a common method of maximizing search sensitivity given a fixed computing budget.
The work of Wette and Prix [Phys.~Rev.~D \textbf{88}, 123005 (2013)] and Wette [Phys.~Rev.~D \textbf{92}, 082003 (2015)] developed a semicoherent search method which uses metrics to construct the banks of pulsar signal templates needed to search the parameter space of interest.
In this work we extend the range of validity of the parameter-space metrics using an empirically-derived relationship between the resolution (or mismatch) of the template banks and the mismatch of the overall search.
This work has important consequences for the optimization of metric-based semicoherent searches at fixed computing cost.
\end{abstract}

\pacs{04.80.Nn, 95.55.Ym, 95.75.Pq, 97.60.Jd}

\maketitle

\section{Introduction}\label{sec:introduction}

The pursuit of the first direct detection of gravitational waves ended with the observation of the merger of two binary black holes~\cite{Abbott.etal.2016a}.
Other classes of gravitational-wave sources (see e.g.~\cite{Pitkin.etal.2011a,Riles.2013a,LSC.VC.2014a} for reviews) may also be detected by the LIGO~\cite{Abbott.etal.2009d,Aasi.etal.2015c}, Virgo~\cite{Acernese.etal.2015a}, and KAGRA~\cite{Somiya.2012a} observatories, as construction and commissioning of these detectors continues over the coming years.

Rapidly-rotating neutron stars which may be radiating continuous, quasisinusoidal gravitational waves -- \emph{gravitational-wave pulsars}, for short -- are one potential source.
Data from the LIGO and Virgo observatories has been searched for gravitational waves from known electromagnetic pulsars~\cite[e.g.][]{Aasi.etal.2014b,Aasi.etal.2015a} and the low-mass X-ray binary Scorpius X-1~\cite[e.g.][]{Aasi.etal.2015b}, gravitational-wave pulsars in supernova remnants~\cite{Abadie.etal.2010b,Aasi.etal.2015f} and at the Galactic center~\cite{Aasi.etal.2013b}, and all-sky searches for gravitational-wave pulsars, both isolated~\cite[e.g.][]{Aasi.etal.2013a,Aasi.etal.2014g,Aasi.etal.2014c,Aasi.etal.2016a} and in binary systems~\cite{Aasi.etal.2014d}.

The detection of gravitational-wave pulsars presents a number of challenges.
The gravitational wave amplitude scales with the potential nonaxial deformability of a neutron star; while \emph{maximum} deformations have been studied e.g.\ in~\cite{JohnsonMcDaniel.Owen.2013a}, the scale of realistic deformations that might exist in the population of Galactic neutron stars, and hence the number of detectable gravitational-wave pulsars, remain largely unknown. Furthermore, a search for gravitational-wave pulsars using the most sensitive search method -- coherent matched filtering against a known signal template -- is computationally feasible only in a few circumstances, e.g.\ searches targeting electromagnetic pulsars whose sky position and frequency evolution are accurately known.

These challenges have motivated the development of a variety of data-analysis techniques, in a quest to gain maximum sensitivity within computational constraints.
These techniques include optimized signal template banks~\cite[e.g.][]{Messenger.etal.2009a,Wette.2014a,Pisarski.Jaranowski.2015a}, semicoherent search methods which trade sensitivity for reduced computing cost~\cite[e.g.][]{Brady.Creighton.2000a,Krishnan.etal.2004a,Dergachev.2010b,Pletsch.2010a,Messenger.2011a,Wette.2015a}, and follow-up procedures for potentially interesting candidate signals~\cite[e.g.][]{Shaltev.Prix.2013a,Behnke.etal.2015a}.

This paper continues a series of papers~\cite{Wette.Prix.2013a,Wette.2014a,Wette.2015a} which have developed a semicoherent search method for isolated gravitational-wave pulsars.
In common with other semicoherent methods, the input gravitational-wave data are partitioned in time into a number of \emph{segments}, each of which is searched by coherent matched filtering against a \emph{coherent} template bank for each segment.
Detection statistics from each segment are then summed together using a distinct \emph{semicoherent} template bank for the overall search.
The method utilizes the idea of a parameter-space \emph{metric}~\cite{Balasubramanian.etal.1996a,Owen.1996a,Prix.2007a}, which determines both the resolution of the coherent template banks of each segment, and that of the semicoherent template bank of the overall search.

The resolutions of the template banks are typically quantified by the maximum \emph{mismatch}: the fraction of signal-to-noise ratio lost when a signal in the input data does not precisely match any one of the search templates.
The parameter-space metric models the mismatch as a distance measure between the parameters of the signal and that of a template.
Typically, the resolution of the semicoherent bank is much finer than that of the coherent banks; in particular the method developed in~\cite{Wette.2015a} predicts that the semicoherent bank requires a much larger number of templates than previously estimated~\cite{Brady.Creighton.2000a,Pletsch.2010a}.

Indeed, the large number of templates in the semicoherent bank leads to the following problem.
The parameters describing the search setup -- the number of segments, the time span of each segment, and the maximum mismatches allowed in the coherent and semicoherent template banks -- may be optimized under fixed constraints on computing cost and available input data using the framework of~\cite{Prix.Shaltev.2012a}.
Preliminary studies have found that, under computing cost and data constraints similar to previous searches~\cite[e.g.][]{Aasi.etal.2013a} performed on the distributed computing project Einstein@Home~\footnote{ \url{http://www.einsteinathome.org/} }, the number of templates in the semicoherent bank becomes very large -- typically $10^{6}$--$10^{7}$ times the number of templates in a coherent bank -- and the computing cost of the semicoherent summation of coherent templates exceeds the computing cost constraint by several orders of magnitude.
In order to reduce the computing cost to that of its constraint, the maximum mismatch allowed in the semicoherent bank must be increased, reducing the number of templates and thereby the computing cost of the semicoherent stage.

The parameter-space metric is, however, only an approximate model based on a Taylor expansion of the mismatch, and hence has a limited range of validity.
Previous work suggests that the metric accurately predicts mismatch values less than $\sim 0.4$ (see e.g.\ Fig.~10 of~\cite{Prix.2007a} and Fig.~7 of~\cite{Wette.Prix.2013a}), and becomes increasingly inaccurate at higher values.
In order to satisfy the computing cost constraint in the search optimization described above, however, the maximum mismatch allowed in the semicoherent bank must typically be \emph{much greater} than $\sim 0.4$, i.e.\ beyond the range of validity of the metric.

It would appear therefore that, under reasonable computing cost constraints, the parameter-space metric \emph{alone} cannot be used to reliably predict mismatch; independent investigations into the performance of Einstein@Home all-sky searches have reached a similar conclusion~\cite{Papa.Walsh.2015}.
Note too that the sensitivity of a search is generally degraded as the maximum allowed mismatch is increased; it is therefore unclear whether, at the large maximum semicoherent mismatch required to satisfy computing cost constraints, the sensitivity of a metric-based search as proposed by~\cite{Wette.Prix.2013a,Wette.2014a,Wette.2015a} would be competitive with other semicoherent methods.

This situation motivates the work described in this paper: a study of the relationship between the mismatches of the coherent and semicoherent template banks predicted by the metric, and the actual mismatch of the overall search as measured by searching for software-generated signals in simulated data.
After reviewing background information in Section~\ref{sec:background}, and the methodology of the simulations used to measure actual mismatch in Section~\ref{sec:numer-simul-at}, the results of the study are presented in Sections~\ref{sec:mean-calf-statistic} and~\ref{sec:calf-stat-mism}.
The conclusions drawn from the study are presented in Section~\ref{sec:discussion}.

\section{Background}\label{sec:background}

This section reviews the theory of semicoherent gravitational-wave pulsar searches, and the associated parameter-space metrics.
Further details can be found in~\cite{Wette.Prix.2013a,Wette.2015a} and references therein.

The signal template $h(t, \vec\calA, \cohsvec\lambda)$ for a gravitational-wave pulsar~\cite{Jaranowski.etal.1998a} is a function of time $t$ at the detector, the four parameters $\vec\calA$ which determine the amplitude modulation of the signal, and the vector of parameters $\cohsvec\lambda$ which determine its phase evolution.
The latter parameters, in the case of all-sky searches for isolated pulsars, are the sky position of the pulsar, its initial frequency $f \equiv \ndot[0] f$ at some reference time $t_0$, and its spindown parameters $\ndot f \equiv d^sf / dt^s |_{t=t_0}$ of spindown order $s$.

The $\calF$-statistic~\cite{Jaranowski.etal.1998a,Cutler.Schutz.2005a} performs matched filtering of the input gravitational-wave data against the template $h(t, \vec\calA, \cohsvec\lambda)$, and analytically maximized over the parameters $\vec\calA$.
If the $\cohsvec\lambda$ are unknown (as is the case for all-sky searches), a search is performed by computing the detection statistic $2\calF\useg(\cohsvec\lambda)$ over a bank of templates whose parameters $\{\cohsvec\lambda\} \in \calP$ are drawn from the search parameter space $\calP$ of interest.
(Here the subscript $\ell$ indexes a single data segment, whose time span is denoted $\coh T$.)
In the vicinity of a signal with parameters $\sigvec\lambda$, the value of $2\calF\useg(\cohsvec\lambda)$ follows a noncentral $\chi^2$ distribution with 4 degrees of freedom and noncentrality parameter $\rho\useg^2(\vec\calA, \sigvec\lambda; \cohsvec\lambda)$; in Gaussian noise it follows a central $\chi^2$ distribution with 4 degrees of freedom.

The mismatch $\cohs\mu\Fstat(\vec\calA, \sigvec\lambda; \cohsvec\lambda)$ determines what fraction of the signal with parameters $\sigvec\lambda$ is not recovered when computing the $\calF$-statistic using a template with parameters $\cohsvec\lambda$.
It is defined in terms of the noncentrality parameter by~\cite[e.g.][]{Prix.Shaltev.2012a,Wette.2015a}:
\begin{equation}
\label{eq:coh-Fstat-mismatch-def}
\cohs\mu\Fstat(\vec\calA, \sigvec\lambda; \cohsvec\lambda) \equiv 1 - \frac{ \rho\useg^2(\vec\calA, \sigvec\lambda; \cohsvec\lambda) }{ \rho\useg^2(\vec\calA, \sigvec\lambda; \sigvec\lambda) } \,,
\end{equation}
where $\rho\useg^2(\vec\calA, \sigvec\lambda; \sigvec\lambda)$ is the noncentrality parameter when the template is perfectly matched to the signal.
A degree of mismatch is unavoidable, as a signal will never exactly match any template in the bank.
Consequentially, template banks are constructed~\cite[e.g.][]{Owen.1996a,Prix.2007b,Wette.2014a} so as to minimize the potential mismatch to some maximum $\coh\mu\umax$.

Mismatch can be modeled~\cite{Balasubramanian.etal.1996a,Owen.1996a,Prix.2007a} as the distance between signal $\sigvec\lambda$ and template $\cohsvec\lambda$ parameters with respect to the parameter-space metric $\cohsmat g\Fstat(\vec\calA, \sigvec\lambda)$:
\begin{equation}
\label{eq:coh-Fstat-mismatch-metric}
\cohs\mu\Fstat(\vec\calA, \sigvec\lambda; \cohsvec\lambda) \approx \sig\diff\cohsvec\lambda \cdot \cohsmat g\Fstat(\vec\calA, \sigvec\lambda) \cdot \sig\diff\cohsvec\lambda \,.
\end{equation}
The metric arises from a second-order Taylor expansion of Eq.~\eqref{eq:coh-Fstat-mismatch-def} with respect to small parameter offsets $\sig\diff\cohsvec\lambda \equiv \cohsvec\lambda - \sigvec\lambda$:
\begin{equation}
\label{eq:coh-Fstat-metric-def}
\cohsmat g\Fstat(\vec\calA, \sigvec\lambda) \equiv \frac{-1}{2 \rho\useg^2(\vec\calA, \sigvec\lambda; \sigvec\lambda)} \left. \frac{\partial \rho\useg^2(\vec\calA, \sigvec\lambda; \cohsvec\lambda)}{\partial \cohsvec\lambda} \right|_{\cohsvec\lambda = \sigvec\lambda} \,.
\end{equation}
Note that whereas by construction the actual mismatch of Eq.~\eqref{eq:coh-Fstat-mismatch-def} may never exceed 1.0, the mismatch predicted by Eq.~\eqref{eq:coh-Fstat-mismatch-metric} may become arbitrarily large as $\| \sig\diff\cohsvec\lambda \|$ increases.

The full metric of Eq.~\eqref{eq:coh-Fstat-metric-def} is often approximated by a simpler expression, the phase metric $\cohsmat g\phase(\sigvec\lambda)$, which discards the amplitude modulation parameterized by the $\vec\calA$.
The components of the matrix $\cohsmat g\phase(\sigvec\lambda)$ are
\begin{equation}
\label{eq:coh-phase-metric-def}
\cohsmat g\phase(\sig\lambda_{ij}) \equiv \big\langle \partial_{\lambda_{i}}\phi \, \partial_{\lambda_{j}}\phi \big\rangle_t - \big\langle \partial_{\lambda_{i}}\phi \big\rangle_t \big\langle \partial_{\lambda_{j}}\phi \big\rangle_t |_{\vec\lambda = \sigvec\lambda} \,,
\end{equation}
where $\phi(t,\vec\lambda)$ is the signal phase, $\partial_{\lambda_{i}}$ are the derivatives with respect to the $i$th parameter in $\vec\lambda$, and the $\langle\cdot\rangle_t$ operator denotes time-averaging over the time span $\coh T$.

It follows from Eq.~\eqref{eq:coh-phase-metric-def} that, for any parameters in $\sigvec\lambda$ in which the signal phase $\phi(t,\sigvec\lambda)$ is linear, the corresponding components of $\cohsmat g\phase(\sigvec\lambda)$ will be independent of $\sigvec\lambda$; the mismatch with respect to a signal $\sigvec\lambda$ will therefore not depend on the parameters of that signal.
This is particularly convenient for template placement since, for such a metric, template banks can be constructed using regular lattices that minimize the number of templates, and hence the number of matched filtering operations~\cite[e.g.][]{Jaranowski.Krolak.2005a,Prix.2007b,Wette.2014a}.
The signal phase $\phi(t,\sigvec\lambda)$ is linear in the frequency $f$ and spindown $\ndot f$ parameters~\cite{Jaranowski.etal.1998a}, but not in commonly-used parameterizations of the sky, e.g.\ by right ascension $\alpha$ and declination $\delta$.

The work of~\cite{Wette.Prix.2013a} developed an approximation to $\cohsmat g\phase(\sigvec\lambda)$ which is independent of all parameters $\sigvec\lambda$: the supersky metric $\cohsmat g$.
The metric is derived by embedding $\cohsmat g\phase(\sigvec\lambda)$ in a higher-dimensional space which includes an additional sky position parameter; in this space $\phi(t,\cohsvec\lambda)$ is linear in all parameters.
Then, the vector in the now 3-dimensional sky parameter space is identified, along which the mismatch $\cohs\mu = \sig\diff\cohsvec\lambda \cdot \cohsmat g \cdot \sig\diff\cohsvec\lambda$ is the least sensitive to the parameter offsets $\sig\diff\cohsvec\lambda$; this vector is the eigenvector associated with the smallest eigenvalue of the sky--sky components of the embedded phase metric.
Finally, the embedded phase metric is projected back onto a subspace perpendicular to this vector, which removes one sky position parameter and results in the metric $\cohsmat g$.
Numerical simulations~\cite{Wette.Prix.2013a} found that $\cohsmat g$ generally predicts mismatches measured by searching for software-generated signals in simulated data with a relative error $\lesssim 30$\%, up to maximum mismatches of $\sim 0.6$.

The metric $\cohsmat g$ applies only to a fully-coherent analysis of a single data segment.
In~\cite{Wette.2015a}, the supersky metric is generalized to a semicoherent analysis, in which the coherent analyses of $N$ data segments are combined together.
For each semicoherent template $\semivec\lambda$ in the bank $\{\semivec\lambda\} \in \calP$, appropriate $\calF$-statistic values $2\calF\useg(\cohsvec\lambda)$ are chosen from each segment and summed to give the $\calF$-statistic as a function of $\semivec\lambda$:
\begin{equation}
\label{eq:semi-Fstat-def}
2\calF\big(\semivec\lambda\big) \equiv \sums 2\calF\useg\big(\cohsvec\lambda\big) \,.
\end{equation}
Typically the $2\calF\useg(\cohsvec\lambda)$ are chosen by nearest-neighbor interpolation: in each segment, the chosen $\cohsvec\lambda$ are the parameters with the smallest mismatch to $\semivec\lambda$ with respect to the metric $\cohsmat g$.

The semicoherent supersky metric $\semimat g$ is used to construct the semicoherent template bank $\{\semivec\lambda\}$, just as the coherent metrics $\cohsmat g$ are used to construct the coherent template banks $\{\cohsvec\lambda\}$ in each segment.
The metric $\semimat g$ is derived following a similar procedure to that of $\cohsmat g$; the chief difference is that its starting point is the phase metric summed over segments $\semimat g\phase(\sigvec\lambda) = \sums \cohsmat g\phase(\sigvec\lambda) / N$~\cite{Brady.Creighton.2000a}.
Numerical simulations using a range of search setups -- parameterized by the number of segments $N$, the time span $\coh T$ of each segment, the total time $\semi T$ spanned by all segments, and the maximum mismatches $\coh\mu\umax$ and $\semi\mu\umax$ of the coherent and semicoherent template banks respectively -- found $\semimat g$ to also be a useful predictor of actual mismatch, with relative errors typically $\lesssim 35$\%, up to maximum mismatches of $\sim 0.5$~\cite{Wette.2015a}.

\section{Numerical simulations at large metric mismatches}\label{sec:numer-simul-at}

\begin{table}
\begin{tabular*}{\linewidth}{r@{\extracolsep{0pt}}l|@{\extracolsep{\fill}}r@{\extracolsep{0pt}}l@{\extracolsep{\fill}}r@{\extracolsep{0pt}}l@{\extracolsep{\fill}}r@{\extracolsep{0pt}}l@{\extracolsep{\fill}}r@{\extracolsep{0pt}}l@{\extracolsep{\fill}}r@{\extracolsep{0pt}}l@{\extracolsep{\fill}}r@{\extracolsep{0pt}}l@{\extracolsep{\fill}}r@{\extracolsep{0pt}}l@{\extracolsep{\fill}}r@{\extracolsep{0pt}}l@{\extracolsep{\fill}}r@{\extracolsep{0pt}}l@{\extracolsep{\fill}}r@{\extracolsep{0pt}}l@{\extracolsep{\fill}}r@{\extracolsep{0pt}}l@{\extracolsep{\fill}}r@{\extracolsep{0pt}}l@{\extracolsep{\fill}}r@{\extracolsep{0pt}}l@{\extracolsep{2\tabcolsep}}}
\hline\hline
\multicolumn{2}{c}{$\semi\mu\umax$} & \multicolumn{26}{c}{$\coh\mu\umax$} \\
\hline
$0$ & $.1$ & $0$ & $.1^{*}$ & \multicolumn{2}{c}{ } & $0$ & $.5^{*}$ & \multicolumn{2}{c}{ } & $1$ & $.5$ & \multicolumn{2}{c}{ } & $4$ & $.1$ & \multicolumn{2}{c}{ } & $10$ & $.9$ & \multicolumn{2}{c}{ } & $28$ & $.7$ & \multicolumn{2}{c}{ } & $75$ & $.3$ \\
$0$ & $.3$ & \multicolumn{2}{c}{ } & $0$ & $.3^{*}$ & \multicolumn{2}{c}{ } & $0$ & $.9$ & \multicolumn{2}{c}{ } & $2$ & $.5$ & \multicolumn{2}{c}{ } & $6$ & $.7$ & \multicolumn{2}{c}{ } & $17$ & $.7$ & \multicolumn{2}{c}{ } & $46$ & $.5$ & \multicolumn{2}{c}{ } \\
$0$ & $.5$ & $0$ & $.1^{*}$ & \multicolumn{2}{c}{ } & $0$ & $.5^{*}$ & \multicolumn{2}{c}{ } & $1$ & $.5$ & \multicolumn{2}{c}{ } & $4$ & $.1$ & \multicolumn{2}{c}{ } & $10$ & $.9$ & \multicolumn{2}{c}{ } & $28$ & $.7$ & \multicolumn{2}{c}{ } & $75$ & $.3$ \\
$0$ & $.9$ & \multicolumn{2}{c}{ } & $0$ & $.3$ & \multicolumn{2}{c}{ } & $0$ & $.9$ & \multicolumn{2}{c}{ } & $2$ & $.5$ & \multicolumn{2}{c}{ } & $6$ & $.7$ & \multicolumn{2}{c}{ } & $17$ & $.7$ & \multicolumn{2}{c}{ } & $46$ & $.5$ & \multicolumn{2}{c}{ } \\
$1$ & $.5$ & $0$ & $.1$ & \multicolumn{2}{c}{ } & $0$ & $.5$ & \multicolumn{2}{c}{ } & $1$ & $.5$ & \multicolumn{2}{c}{ } & $4$ & $.1$ & \multicolumn{2}{c}{ } & \multicolumn{2}{c}{ } & \multicolumn{2}{c}{ } & \multicolumn{2}{c}{ } & \multicolumn{2}{c}{ } & \multicolumn{2}{c}{ } \\
$2$ & $.5$ & \multicolumn{2}{c}{ } & $0$ & $.3$ & \multicolumn{2}{c}{ } & $0$ & $.9$ & \multicolumn{2}{c}{ } & $2$ & $.5$ & \multicolumn{2}{c}{ } & \multicolumn{2}{c}{ } & \multicolumn{2}{c}{ } & \multicolumn{2}{c}{ } & \multicolumn{2}{c}{ } & \multicolumn{2}{c}{ } & \multicolumn{2}{c}{ } \\
$4$ & $.1$ & $0$ & $.1$ & \multicolumn{2}{c}{ } & $0$ & $.5$ & \multicolumn{2}{c}{ } & $1$ & $.5$ & \multicolumn{2}{c}{ } & \multicolumn{2}{c}{ } & \multicolumn{2}{c}{ } & \multicolumn{2}{c}{ } & \multicolumn{2}{c}{ } & \multicolumn{2}{c}{ } & \multicolumn{2}{c}{ } & \multicolumn{2}{c}{ } \\
$6$ & $.7$ & \multicolumn{2}{c}{ } & $0$ & $.3$ & \multicolumn{2}{c}{ } & $0$ & $.9$ & \multicolumn{2}{c}{ } & \multicolumn{2}{c}{ } & \multicolumn{2}{c}{ } & \multicolumn{2}{c}{ } & \multicolumn{2}{c}{ } & \multicolumn{2}{c}{ } & \multicolumn{2}{c}{ } & \multicolumn{2}{c}{ } & \multicolumn{2}{c}{ } \\
$10$ & $.9$ & $0$ & $.1$ & \multicolumn{2}{c}{ } & $0$ & $.5$ & \multicolumn{2}{c}{ } & \multicolumn{2}{c}{ } & \multicolumn{2}{c}{ } & \multicolumn{2}{c}{ } & \multicolumn{2}{c}{ } & \multicolumn{2}{c}{ } & \multicolumn{2}{c}{ } & \multicolumn{2}{c}{ } & \multicolumn{2}{c}{ } & \multicolumn{2}{c}{ } \\
$17$ & $.7$ & \multicolumn{2}{c}{ } & $0$ & $.3$ & \multicolumn{2}{c}{ } & $0$ & $.9$ & \multicolumn{2}{c}{ } & \multicolumn{2}{c}{ } & \multicolumn{2}{c}{ } & \multicolumn{2}{c}{ } & \multicolumn{2}{c}{ } & \multicolumn{2}{c}{ } & \multicolumn{2}{c}{ } & \multicolumn{2}{c}{ } & \multicolumn{2}{c}{ } \\
$28$ & $.7$ & $0$ & $.1$ & \multicolumn{2}{c}{ } & $0$ & $.5$ & \multicolumn{2}{c}{ } & \multicolumn{2}{c}{ } & \multicolumn{2}{c}{ } & \multicolumn{2}{c}{ } & \multicolumn{2}{c}{ } & \multicolumn{2}{c}{ } & \multicolumn{2}{c}{ } & \multicolumn{2}{c}{ } & \multicolumn{2}{c}{ } & \multicolumn{2}{c}{ } \\
$46$ & $.5$ & \multicolumn{2}{c}{ } & $0$ & $.3$ & \multicolumn{2}{c}{ } & $0$ & $.9$ & \multicolumn{2}{c}{ } & \multicolumn{2}{c}{ } & \multicolumn{2}{c}{ } & \multicolumn{2}{c}{ } & \multicolumn{2}{c}{ } & \multicolumn{2}{c}{ } & \multicolumn{2}{c}{ } & \multicolumn{2}{c}{ } & \multicolumn{2}{c}{ } \\
$75$ & $.3$ & $0$ & $.1$ & \multicolumn{2}{c}{ } & $0$ & $.5$ & \multicolumn{2}{c}{ } & \multicolumn{2}{c}{ } & \multicolumn{2}{c}{ } & \multicolumn{2}{c}{ } & \multicolumn{2}{c}{ } & \multicolumn{2}{c}{ } & \multicolumn{2}{c}{ } & \multicolumn{2}{c}{ } & \multicolumn{2}{c}{ } & \multicolumn{2}{c}{ } \\
\hline
\end{tabular*}
\caption{\label{tab:max_mismatch}
Pairs of maximum semicoherent and coherent template bank mismatches $(\semi\mu\umax, \coh\mu\umax)$ used by the numerical simulations described in Section~\ref{sec:numer-simul-at}.
For a given row, which fixes $\semi\mu\umax$ to the value given in the first column, the presence of values of $\coh\mu\umax$ in subsequent columns indicates the $(\semi\mu\umax, \coh\mu\umax)$ pairs that were used.
The $(\semi\mu\umax, \coh\mu\umax)$ pairs used by the numerical simulations in~\cite{Wette.2015a} are asterisked.
}
\end{table}

In~\cite{Wette.Prix.2013a,Wette.2015a} the supersky metric $\semimat g$ was validated by performing numerical simulations which compare the mismatch predictions of the metric to mismatches measured by searching simulated data for software-generated signals.
Those simulations limited the maximum mismatches of both the coherent and semicoherent template banks to $\le 0.6$~\cite{Wette.Prix.2013a} and $\le 0.5$~\cite{Wette.2015a}.
In this paper we reperform the same simulations with a much wider range of maximum mismatches $\le 75.3$.
The simulation procedure, which is otherwise very similar to that used in~\cite{Wette.Prix.2013a,Wette.2015a}, is briefly described in this section.

A total of 48 pairs of maximum semicoherent and coherent mismatches $(\semi\mu\umax, \coh\mu\umax)$ are used; these are listed in Table~\ref{tab:max_mismatch}.
The simulations used a variety of search setups, parameterized by $(\semi T, \coh T, \eta)$, where $\eta = N \coh T / \semi T$ is the segment duty cycle, i.e.\ the fraction of the total time span $\semi T$ which falls within a segment.
The chosen parameters are $\semi T \in \{120, 240, 360\}$~days, $\coh T \in \{1, 3, 5, 7\}$~days, and $\eta \in \{50\%, 75\%, 100\%\}$; these are a subset of the parameters used in~\cite{Wette.2015a}.

Signal parameters $\sigvec\lambda = (\alpha, \delta, f, \ndot f)$ are generated with: uniform sky positions; spindowns such that $|\ndot f| \le 0.5 \beta_{\ndot f}$, where $\beta_{\ndot f}$ is the width of the metric ellipse bounding box of $\semimat g$, and is given by Eq.~(11) of~\cite{Wette.2014a}; and frequencies $f \in \{100, 1000\}$~Hz.
The nearest semicoherent template $\semivec\lambda$ to the signal is determined assuming a semicoherent template bank constructed from an $\Ans[4]$ lattice~\cite{Conway.Sloane.1988} using the metric $\semimat g$ with maximum mismatch $\semi\mu\umax$ chosen from one of the pairs in Table~\ref{tab:max_mismatch}.
Similarly, in each segment the nearest coherent template $\cohsvec\lambda$ to the semicoherent template is determined assuming a coherent template bank constructed from an $\Ans[4]$ lattice using the metric $\cohsmat g$ with maximum mismatch $\coh\mu\umax$ chosen from the same pair in Table~\ref{tab:max_mismatch}.

Metric mismatches between the signal $\sigvec\lambda$ and various templates are then calculated~\footnote{
Note that this paper uses a slightly different notation than~\cite{Prix.Shaltev.2012a,Wette.2015a} to distinguish between the various metric mismatches.
The total metric mismatch, including contributions from both semicoherent and coherent template banks, is denoted $\mu$ in this paper, as opposed to $\protect\widehat{\mu}$ in~\cite{Prix.Shaltev.2012a,Wette.2015a}.
The component of the total metric mismatch contributed by the semicoherent template bank only is denoted $\protect\semi\mu$ in this paper, as opposed to $\protect\widehat{\mu}_0$ in~\cite{Prix.Shaltev.2012a} and $\protect\widehat{\mu}_{\textrm{ni}}$ in~\cite{Wette.2015a}.
The component of the total metric mismatch contributed by each coherent template bank only is denoted $\protect\coh\mu$ in this paper, in common with~\cite{Prix.Shaltev.2012a,Wette.2015a}.
}. The total metric mismatch between $\sigvec\lambda$ and the nearest $\cohsvec\lambda$ in the coherent template banks of each segment is
\begin{equation}
\label{eq:total-ssky-mismatch}
\mu \equiv \frac{1}{N} \sums \sig\diff\cohsvec\lambda \cdot \cohsmat g \cdot \sig\diff\cohsvec\lambda \,.
\end{equation}
The semicoherent metric mismatch between $\sigvec\lambda$ and the nearest $\semivec\lambda$ in the semicoherent template bank is
\begin{equation}
\label{eq:semi-ssky-mismatch}
\semi\mu \equiv \sig\diff\semivec\lambda \cdot \semimat g \cdot \sig\diff\semivec\lambda \,,
\end{equation}
with $\sig\diff\semivec\lambda \equiv \semivec\lambda - \sigvec\lambda$.
Finally, the average coherent metric mismatch between $\semivec\lambda$ and the nearest $\cohsvec\lambda$ in the coherent template banks of each segment is
\begin{equation}
\label{eq:avg-coh-ssky-mismatch}
\mean{\coh\mu} \equiv \frac{1}{N} \sums \semi\diff\cohsvec\lambda \cdot \cohsmat g \cdot \semi\diff\cohsvec\lambda \,,
\end{equation}
with $\semi\diff\cohsvec\lambda \equiv \cohsvec\lambda - \semivec\lambda$.

The metric mismatches given above are then measured using the $\calF$-statistic~\footnote{ As implemented in the software package \texttt{LALSuite}, available from \url{https://wiki.ligo.org/DASWG/LALSuite}. }.
Simulated gravitational-wave data from the LIGO Livingston detector~\cite{Abbott.etal.2009d} are generated at times $[t_0 - \semi T/2, t_0 + \semi T/2]$, where $t_0 \equiv~$UTC 2015-01-01 00:00:00.
The data comprise no noise (but see the discussion in Section~\ref{sec:discussion}) and a simulated signal with random amplitude parameters $\calA$ and chosen phase parameters $\sigvec\lambda$.
The $\calF$-statistic is computed in each segment at $\sigvec\lambda$, the nearest $\semivec\lambda$ to $\sigvec\lambda$, and the nearest $\cohsvec\lambda$ to $\semivec\lambda$, and the total $\calF$-statistic mismatch is computed using
\begin{equation}
\label{eq:total-Fstat-mismatch}
\mu\Fstat \equiv \frac{1}{N} \sums \frac{ 2\calF\useg(\sigvec\lambda) - 2\calF\useg(\cohsvec\lambda) }{ 2\calF\useg(\sigvec\lambda) - 4 } \,,
\end{equation}
where the denominator equals the noncentrality parameter $\rho\useg^2(\vec\calA, \sigvec\lambda; \sigvec\lambda)$.
The total $\calF$-statistic mismatch is also computed assuming no nearest-neighbor interpolation, i.e.\ that $\cohsvec\lambda = \semivec\lambda$ in every segment:
\begin{equation}
\label{eq:semi-Fstat-mismatch-no-interp}
\mu\Fstat\unoint \equiv \frac{1}{N} \sums \frac{ 2\calF\useg(\sigvec\lambda) - 2\calF\useg(\semivec\lambda) }{ 2\calF\useg(\sigvec\lambda) - 4 } \,.
\end{equation}

This procedure is repeated $10^{5}$ times for all 3456 combinations of $(\semi\mu\umax, \coh\mu\umax)$, $\semi T$, $\coh T$, $\eta$, and $f$. A total of $7.8 \times 10^{10}$ coherent $\calF$-statistic values were computed.

\section{Mean $\calF$-statistic mismatch}\label{sec:mean-calf-statistic}

\begin{figure*}
\centering
\subfloat[]{\includegraphics[width=\linewidth]{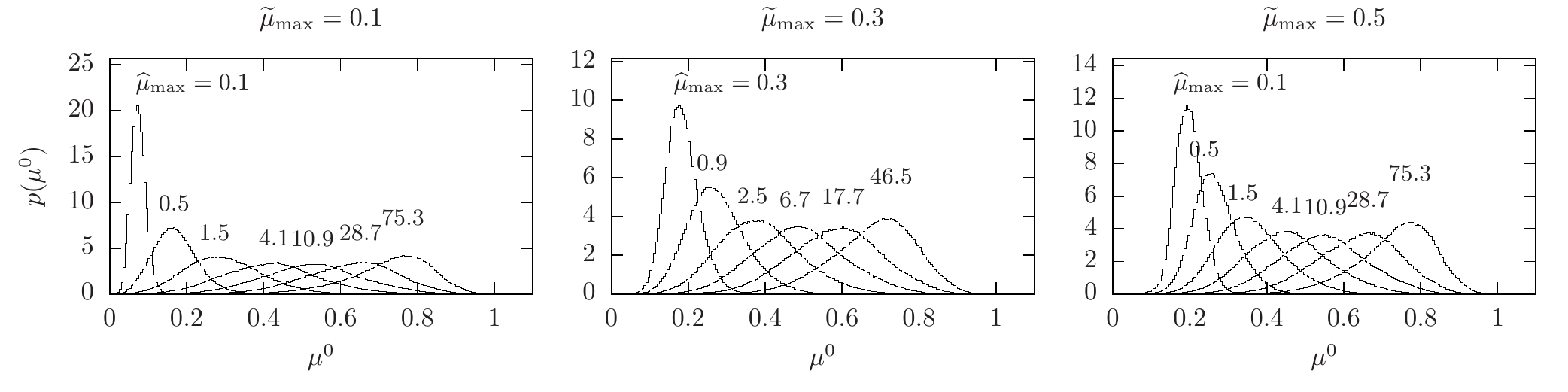}\label{fig:mutwoF_hgrm_Tcoh1d_Tsemi120d}}\\
\subfloat[]{\includegraphics[width=\linewidth]{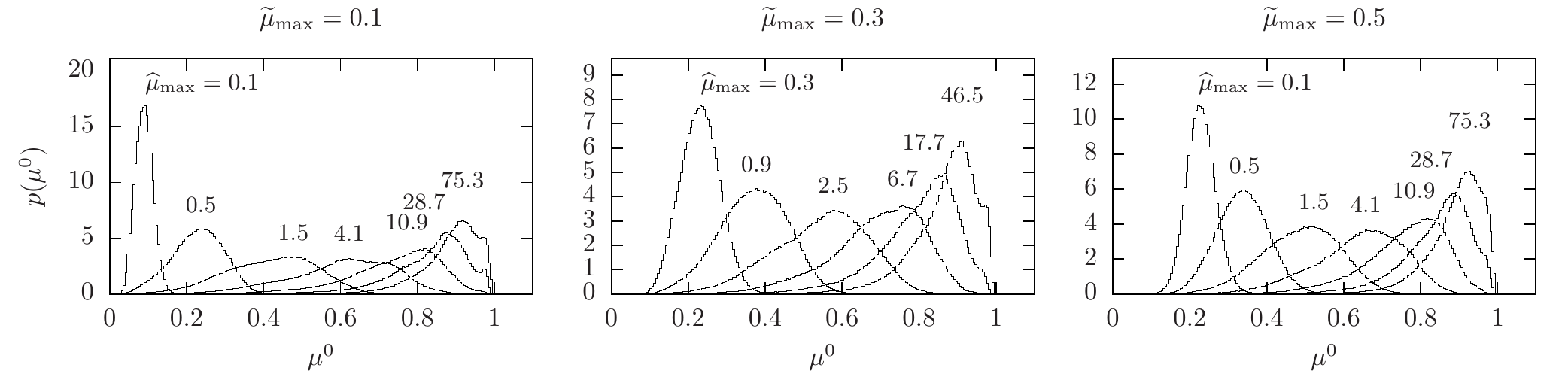}\label{fig:mutwoF_hgrm_Tcoh7d_Tsemi120d}}\\
\subfloat[]{\includegraphics[width=\linewidth]{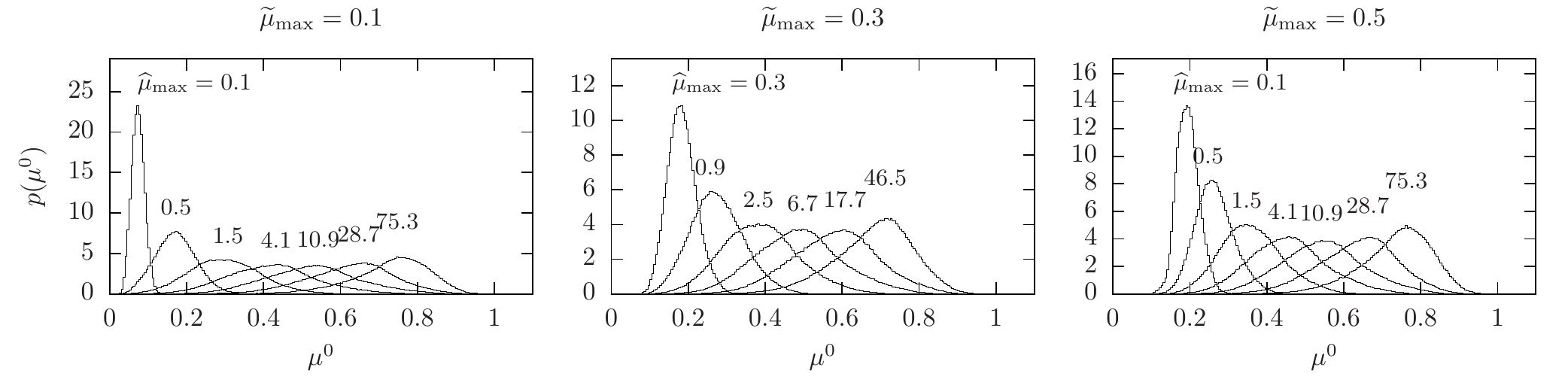}\label{fig:mutwoF_hgrm_Tcoh1d_Tsemi360d}}\\
\subfloat[]{\includegraphics[width=\linewidth]{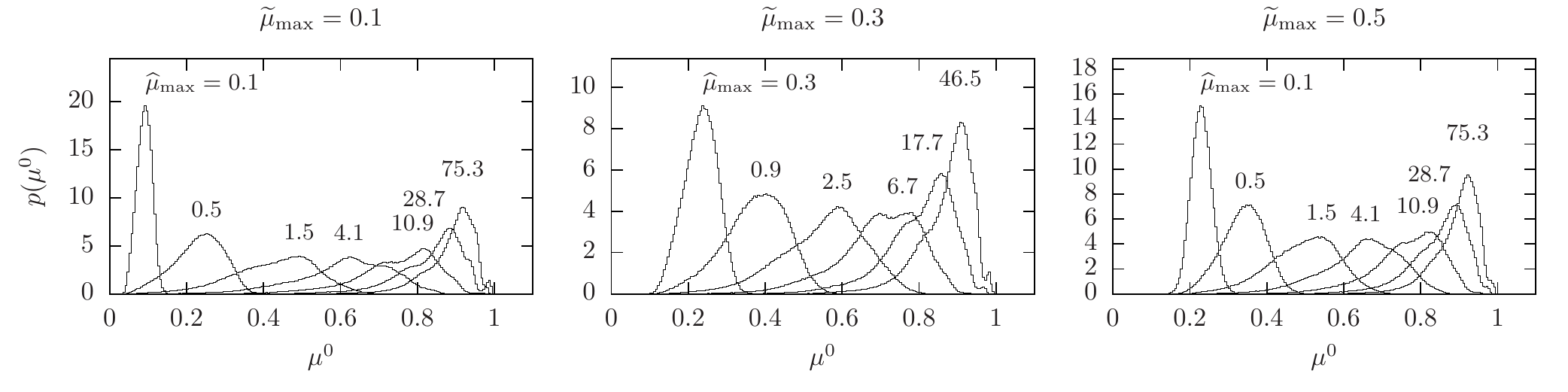}\label{fig:mutwoF_hgrm_Tcoh7d_Tsemi360d}}
\caption{\label{fig:mutwoF_hgrm}
Histograms of $\calF$-statistic mismatch $\mu\Fstat$ for search setups with
\protect\subref{fig:mutwoF_hgrm_Tcoh1d_Tsemi120d} $\coh T = 1$~day, $\semi T = 120$~days,
\protect\subref{fig:mutwoF_hgrm_Tcoh7d_Tsemi120d} $\coh T = 7$~days, $\semi T = 120$~days,
\protect\subref{fig:mutwoF_hgrm_Tcoh1d_Tsemi360d} $\coh T = 1$~day, $\semi T = 360$~days, and
\protect\subref{fig:mutwoF_hgrm_Tcoh7d_Tsemi360d} $\coh T = 7$~days, $\semi T = 360$~days.
The plots are for maximum coherent metric mismatches $\coh\mu\umax$ of 0.1 (left column), 0.3 (middle column), and 0.5 (right column).
Within each plot, the maximum semicoherent metric mismatch $\semi\mu\umax$ of each histogram is labelled.
}
\end{figure*}

\begin{figure*}
\centering
\subfloat[]{\includegraphics[width=0.49\linewidth]{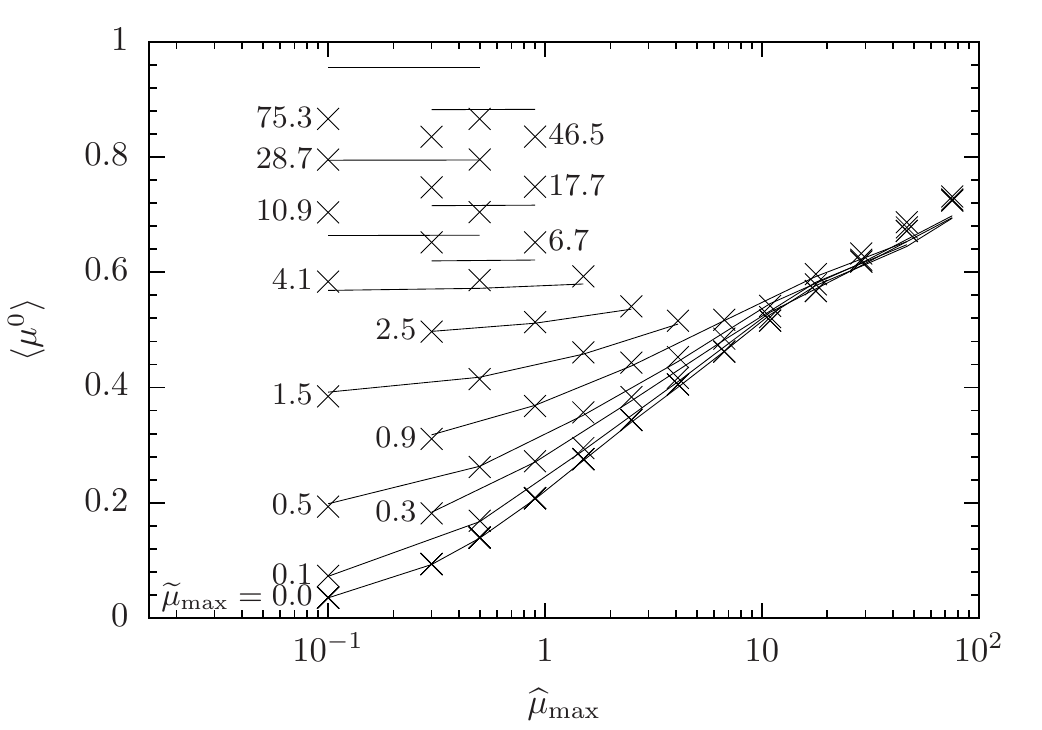}\label{fig:mutwoF_fit_Tcoh1d_Tsemi240d}}
\subfloat[]{\includegraphics[width=0.49\linewidth]{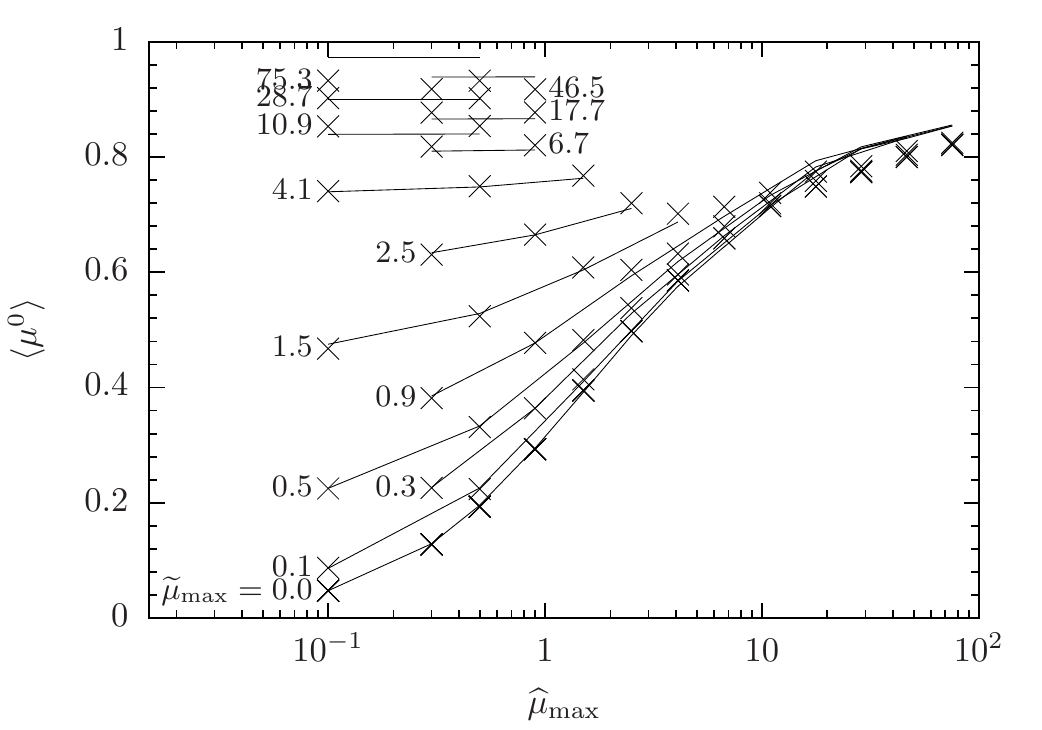}\label{fig:mutwoF_fit_Tcoh3d_Tsemi240d}}\\
\subfloat[]{\includegraphics[width=0.49\linewidth]{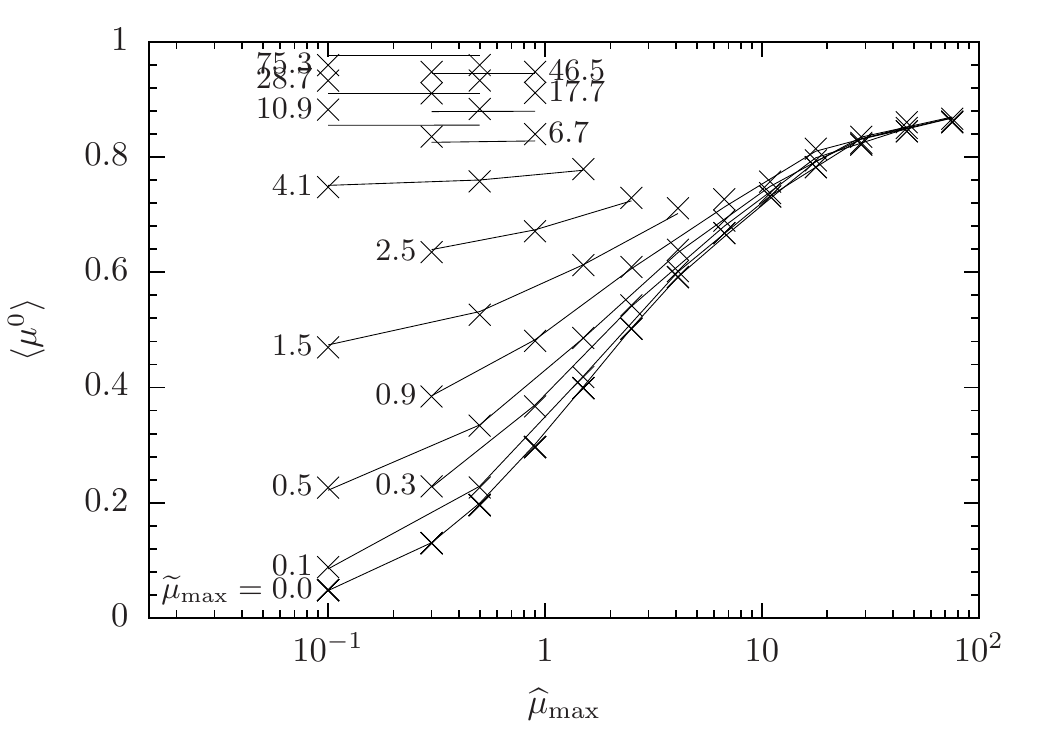}\label{fig:mutwoF_fit_Tcoh5d_Tsemi240d}}
\subfloat[]{\includegraphics[width=0.49\linewidth]{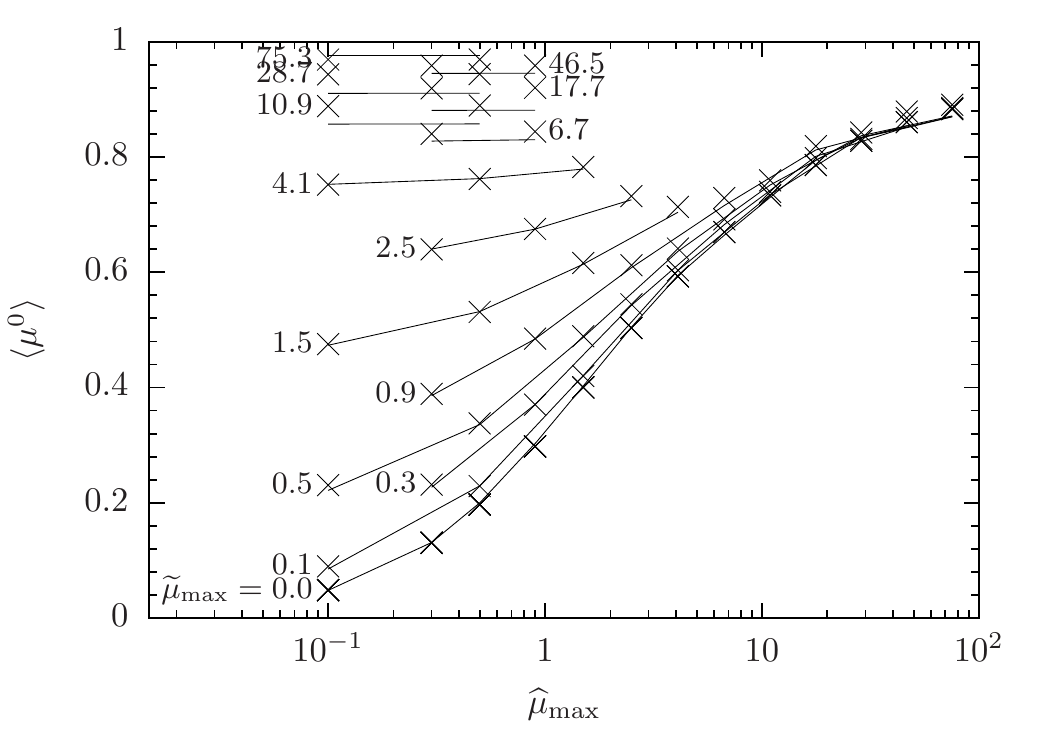}\label{fig:mutwoF_fit_Tcoh7d_Tsemi240d}}
\caption{\label{fig:mutwoF_fit}
Mean $\calF$-statistic mismatch $\mean{\mu\Fstat}$ as a function of $(\semi\mu\umax, \coh\mu\umax)$, for $\semi T = 240$~days and
\protect\subref{fig:mutwoF_fit_Tcoh1d_Tsemi240d} $\coh T = 1$~day,
\protect\subref{fig:mutwoF_fit_Tcoh3d_Tsemi240d} $\coh T = 3$~days
\protect\subref{fig:mutwoF_fit_Tcoh5d_Tsemi240d} $\coh T = 5$~day, and
\protect\subref{fig:mutwoF_fit_Tcoh7d_Tsemi240d} $\coh T = 7$~days.
In each plot, $\mean{\mu\Fstat}$ is plotted as a function of $\semi\mu\umax$ at a fixed value of $\coh\mu\umax$ (crosses), labeled on the left side of each plot.
The value $\coh\mu\umax = 0.0$ denotes no nearest-neighbor interpolation, i.e.\ $\mu\Fstat = \mu\Fstat\unoint$.
For the same $\coh\mu\umax$, the empirical fit to $\mean{\mu\Fstat}$ given by Eq.~\eqref{eq:empr-mean-Fstat-fit} is also plotted (lines).
}
\end{figure*}

\begin{table}
\begin{tabular*}{\linewidth}{l@{\extracolsep{\fill}}r@{\extracolsep{0pt}}l@{\extracolsep{\fill}}r@{\extracolsep{0pt}}l@{\extracolsep{\fill}}r@{\extracolsep{0pt}}l@{\extracolsep{\fill}}r@{\extracolsep{0pt}}l@{\extracolsep{\fill}}r@{\extracolsep{0pt}}l@{\extracolsep{2\tabcolsep}}}
\hline\hline
 & \multicolumn{10}{c}{Coefficients} \\
$n$ & \multicolumn{2}{c}{$1$} & \multicolumn{2}{c}{$2$} & \multicolumn{2}{c}{$3$} & \multicolumn{2}{c}{$4$} & \multicolumn{2}{c}{$5$} \\
\hline
$a\Ufit_{n}$ & $1$ & $.4813$ & $1$ & $.2774$ & $0$ & $.7994$ & $0$ & $.7016$ & $1$ & $.0318$ \\
$b\Ufit_{n}$ & $0$ & $.651$ & $1$ & $.0356$ & $0$ & $.97756$ & $1$ & $.1154$ & $0$ & $.99694$ \\
$c\Ufit_{n}$ & $-1$ & $.7281$ & $-0$ & $.34311$ & $-0$ & $.54179$ & $-0$ & $.41982$ & $-0$ & $.62284$ \\
$d\Ufit_{n}$ & $-1$ & $.3531$ & $-0$ & $.98319$ & $-1$ & $.0903$ & $-0$ & $.98934$ & $-1$ & $.0375$ \\
$e\Ufit_{n}$ & $0$ & $.0028833$ & $1$ & $.3869$ & $0$ & $.093245$ & $0$ & $.47769$ & $0$ & $.74859$ \\
$f\Ufit_{n}$ & $0$ & $.015045$ & $0$ & $.66447$ & $0$ & $.31791$ & $0$ & $.48776$ & $0$ & $.74612$ \\
\hline\hline
\end{tabular*}
\caption{\label{tab:mean_fit_coefficients}
Coefficients of fit for Equation~\eqref{eq:empr-mean-Fstat-fit}.
}
\end{table}

\begin{table}
\begin{tabular*}{\linewidth}{l@{\extracolsep{\fill}}r@{\extracolsep{0pt}}l@{\extracolsep{\fill}}r@{\extracolsep{0pt}}l@{\extracolsep{\fill}}r@{\extracolsep{0pt}}l@{\extracolsep{\fill}}r@{\extracolsep{0pt}}l@{\extracolsep{2\tabcolsep}}}
\hline\hline
 & \multicolumn{4}{c}{$\Zn$ lattice} & \multicolumn{4}{c}{$\Ans$ lattice} \\
$n$ & \multicolumn{2}{c}{$\mean{\mu}/\mu\umax$} & \multicolumn{2}{c}{$\stdv{\mu}/\mu\umax$} & \multicolumn{2}{c}{$\mean{\mu}/\mu\umax$} & \multicolumn{2}{c}{$\stdv{\mu}/\mu\umax$} \\
\hline
$1$ & $0$ & $.33$ & $0$ & $.3$ & $0$ & $.33$ & $0$ & $.3$ \\
$2$ & $0$ & $.33$ & $0$ & $.21$ & $0$ & $.42$ & $0$ & $.24$ \\
$3$ & $0$ & $.33$ & $0$ & $.17$ & $0$ & $.47$ & $0$ & $.22$ \\
$4$ & $0$ & $.33$ & $0$ & $.15$ & $0$ & $.52$ & $0$ & $.2$ \\
$5$ & $0$ & $.33$ & $0$ & $.13$ & $0$ & $.55$ & $0$ & $.18$ \\
\hline\hline
\end{tabular*}
\caption{\label{tab:lattice_mean_stdv_table}
Means $\mean{\mu}$ and standard deviations $\stdv{\mu}$ of the metric mismatch distributions expected from template placement using $\Zn$ and $\Ans$ lattices in $n$ dimensions.
}
\end{table}

In this section, the results of the simulations described in Section~\ref{sec:numer-simul-at} are used to investigate the relationship between predicted metric mismatch and actual $\calF$-statistic mismatch, in the limit of large metric mismatches.

Figure~\ref{fig:mutwoF_hgrm} plots histograms of the mismatches $\mu\Fstat$ measured using the $\calF$-statistic [via Eqs.~\eqref{eq:total-Fstat-mismatch}] as a function of the search setup parameters $\coh T$, $\semi T$, $\coh\mu\umax$, and $\semi\mu\umax$.
As $\semi\mu\umax$ increases, the means of the $\mu\Fstat$ histograms also increase, as expected, but at a slower rate, and do not saturate at 100\% mismatch even when $\semi\mu\umax \gg 1$.
For example, for $\coh T = 1$~day, $\semi T = 120$~days, and $\coh\mu\umax = 0.1$ (left-most subplot in Fig.~\ref{fig:mutwoF_hgrm_Tcoh1d_Tsemi120d}), the mean of the $\mu\Fstat$ histogram for which $\semi\mu\umax = 0.5$ is 0.17; for $\semi\mu\umax = 1.5$, the mean is 0.29; and for $\semi\mu\umax = 10.9$, the mean is 0.53.
This suggests that the number of semicoherent templates could be reduced significantly (by increasing $\semi\mu\umax$) while limiting losses in signal-to-noise ratio (as measured by $\mu\Fstat$) to a reasonable level.

As an example, note that the number of semicoherent templates scales with $\semi\mu\umax^{-n/2}$ where $n = 4$ is the number of parameter-space dimensions (see e.g. Eq.~(29) of~\cite{Wette.2014a}), and that the search sensitivity scales with $\sqrt{1 - \mean{\mu\Fstat}}$~\cite{Wette.2012a}.
By increasing $\semi\mu\umax$ from 0.5 to 10.9, one would save a factor of $(0.5 / 10.9)^{-4/2} \sim 480$ in the number of semicoherent templates.
The corresponding loss in sensitivity from increasing the mean measured mismatch from 0.17 to 0.53 is $1 - \sqrt{ (1 - 0.53)/(1 - 0.17) } \sim 25\%$.
An optimization procedure could potentially recover this loss, however, by reinvesting the computational power saved from the reduction in the number of semicoherent templates, e.g. by increasing $\coh T$; see the discussion in Section~\ref{sec:discussion}.

From Fig.~\ref{fig:mutwoF_hgrm} we see that the mean $\mu\Fstat$ also increases with $\coh\mu\umax$ (compare the left, middle, and right columns), and with $\coh T$ (compare e.g.\ Figs.~\ref{fig:mutwoF_hgrm_Tcoh1d_Tsemi120d} and~\ref{fig:mutwoF_hgrm_Tcoh7d_Tsemi120d}), but is largely independent of $\semi T$ (compare e.g.\ Figs.~\ref{fig:mutwoF_hgrm_Tcoh1d_Tsemi120d} and~\ref{fig:mutwoF_hgrm_Tcoh1d_Tsemi360d}).

Figure~\ref{fig:mutwoF_fit} plots the mean $\calF$-statistic mismatch $\mean{\mu\Fstat}$ as a function of the maximum metric mismatches $(\semi\mu\umax, \coh\mu\umax)$, for $\semi T = 240$~days and the 4 values of $\coh T \in \{1, 3, 5, 7\}$~days.
In keeping with Fig.~\ref{fig:mutwoF_hgrm}, one sees that $\mean{\mu\Fstat}$ increases with $\semi\mu\umax$, but at a slower rate which is roughly a constant per decade in $\semi\mu\umax$.
For example, with $\coh T = 1$~day (Fig.~\ref{fig:mutwoF_fit_Tcoh1d_Tsemi240d}) and $\coh\mu\umax = 0.0$ (i.e.\ $\mu\Fstat = \mu\Fstat\unoint$), $\mean{\mu\Fstat}$ increases from 0.035 to 0.21 as $\semi\mu\umax$ increases from 0.1 to 0.9, i.e.\ at a rate of $\sim 6.5$ per decade in $\semi\mu\umax$; $\mean{\mu\Fstat}$ then increases to 0.52 at $\semi\mu\umax = 10.9$, at a slower rate of $\sim 2$ per decade in $\semi\mu\umax$.
Even at $\semi\mu\umax = 75.3$, $\mean{\mu\Fstat} \in \{ \sim 0.7, \sim 0.9 \}$ has yet to reach 100\% mismatch.

The rate of increase of $\mean{\mu\Fstat}$ is slightly lower at $\coh T = 1$~day (Fig.~\ref{fig:mutwoF_fit_Tcoh1d_Tsemi240d}) than for $\coh T > 1$~day (Figs.~\ref{fig:mutwoF_fit_Tcoh3d_Tsemi240d}--~\ref{fig:mutwoF_fit_Tcoh7d_Tsemi240d}); consequentially, the mean $\calF$-statistic as $\semi\mu\umax$ approaches 100 is lower at $\coh T = 1$~day ($\sim 0.7$ at $\coh\mu\umax = 0.0$) than for $\coh T > 1$~day ($\sim 0.83$).
As $\coh\mu\umax$ is increased, the rate of increase of $\mean{\mu\Fstat}$ with $\semi\mu\umax$ decreases still further, and becomes essentially zero for $\coh\mu\umax \gtrsim 4$.
The behavior of $\mean{\mu\Fstat}$ is only weakly dependent on $\semi T$, and for this reason $\semi T$ is fixed to 240~days in Fig.~\ref{fig:mutwoF_fit}.

We find the following empirical fit to $\mean{\mu\Fstat}$ as a function of $\coh T$, $\semi T$, $\mean{\coh\mu}$, and $\mean{\semi\mu}$:
\begin{equation}
\label{eq:empr-mean-Fstat-fit}
\mean{\mu\Fstat}\Ufit\Big( \coh T, \semi T, \mean{\semi\mu}, \mean{\coh\mu} \Big) = 1 - \frac{
\sum_{n=1}^{5} \exp\big( -y_{n}^{2} \big) / n
}{
\sum_{n=1}^{5} \exp\big( -x_{n}^{2} \big) / n
} \,,
\end{equation}
where
\begin{align}
\label{eq:empr-mean-Fstat-fit-X}
x_{n} &= a\Ufit_{n} + \exp\left[ b\Ufit_{n} + c\Ufit_{n} \, \frac{\semi T}{\text{year}} + d\Ufit_{n} \, \frac{\coh T}{\text{day}} \right] \,, \\
\label{eq:empr-mean-Fstat-fit-Y}
y_{n} &= x_{n} + e\Ufit_{n} \mean{\semi\mu} + f\Ufit_{n} \mean{\coh\mu} \,,
\end{align}
and the fitted coefficients $a\Ufit_{n}$ through $f\Ufit_{n}$ are listed in Table~\ref{tab:mean_fit_coefficients}.
Note that Equation~\eqref{eq:empr-mean-Fstat-fit} uses as parameters the \emph{mean} semicoherent and coherent metric mismatches $\mean{\semi\mu}$ and $\mean{\coh\mu}$, instead of the \emph{maxima} $\semi\mu\umax$ and $\coh\mu\umax$ respectively.
For template banks generated using lattice template placement~\cite{Wette.2014a}, the ratios $\mean{\semi\mu}/\semi\mu\umax$ and $\mean{\coh\mu}/\coh\mu\umax$ are given by the number of parameter-space dimensions and the type of lattice employed; values for up to $n = 5$ dimensions and for $\Zn$ and $\Ans$ lattices are listed in Table~\ref{tab:lattice_mean_stdv_table}.
An empirical fit to the standard deviations $\stdv{\mu\Fstat}$ of the $\calF$-statistic mismatch, i.e.\ the widths of the histograms plotted in Fig.~\ref{fig:mutwoF_hgrm}, is given in Appendix~\ref{sec:stand-devi-calf}.

Over the 576 values of $\mean{\mu\Fstat}$ parameterized by $( \coh T, \semi T, \mean{\semi\mu}, \mean{\coh\mu} )$ used for fitting, the root-mean-square relative error to $\mean{\mu\Fstat}\Ufit$ was minimized to $\lesssim 2$\%.
Each value of $\mean{\mu\Fstat}$ was weighted by the standard deviation of the means of $\mu\Fstat$ as a function of $\eta$ and $f$, the two simulation parameters (see Section~\ref{sec:numer-simul-at}) not included in the fit; these standard deviations are typically $\sim 10^{-3}$.

The empirical fit $\mean{\mu\Fstat}\Ufit$ is plotted~\footnote{ Using an implementation of Eq.~\eqref{eq:empr-mean-Fstat-fit} in the Octave software repository \texttt{OctApps}, available from \url{https://gitlab.aei.uni-hannover.de/octapps/octapps}. } alongside the fitted data $\mean{\mu\Fstat}$ in Fig.~\ref{fig:mutwoF_fit}.
The fit is worst for large $\coh\mu\umax$ at $\coh T = 1$~day, where $\mean{\mu\Fstat}\Ufit$ overestimates $\mean{\mu\Fstat}$ by up to $\sim 15$\%, but improves as $\coh\mu\umax$ decreases and $\coh T$ increases.
The weak dependence of Eq.~\eqref{eq:empr-mean-Fstat-fit} on $\semi T$ can be seen in Eq.~\eqref{eq:empr-mean-Fstat-fit-X}; note that $|b\Ufit|$, $|c\Ufit|$, and $|d\Ufit|$ are of order unity, and that while $\coh T$ would typically be greater than 1~day, $\semi T$ is typically of order 1~year or less.

\section{$\calF$-statistic mismatch as function of search parameters}\label{sec:calf-stat-mism}

\begin{figure*}
\subfloat[]{\includegraphics[width=\linewidth]{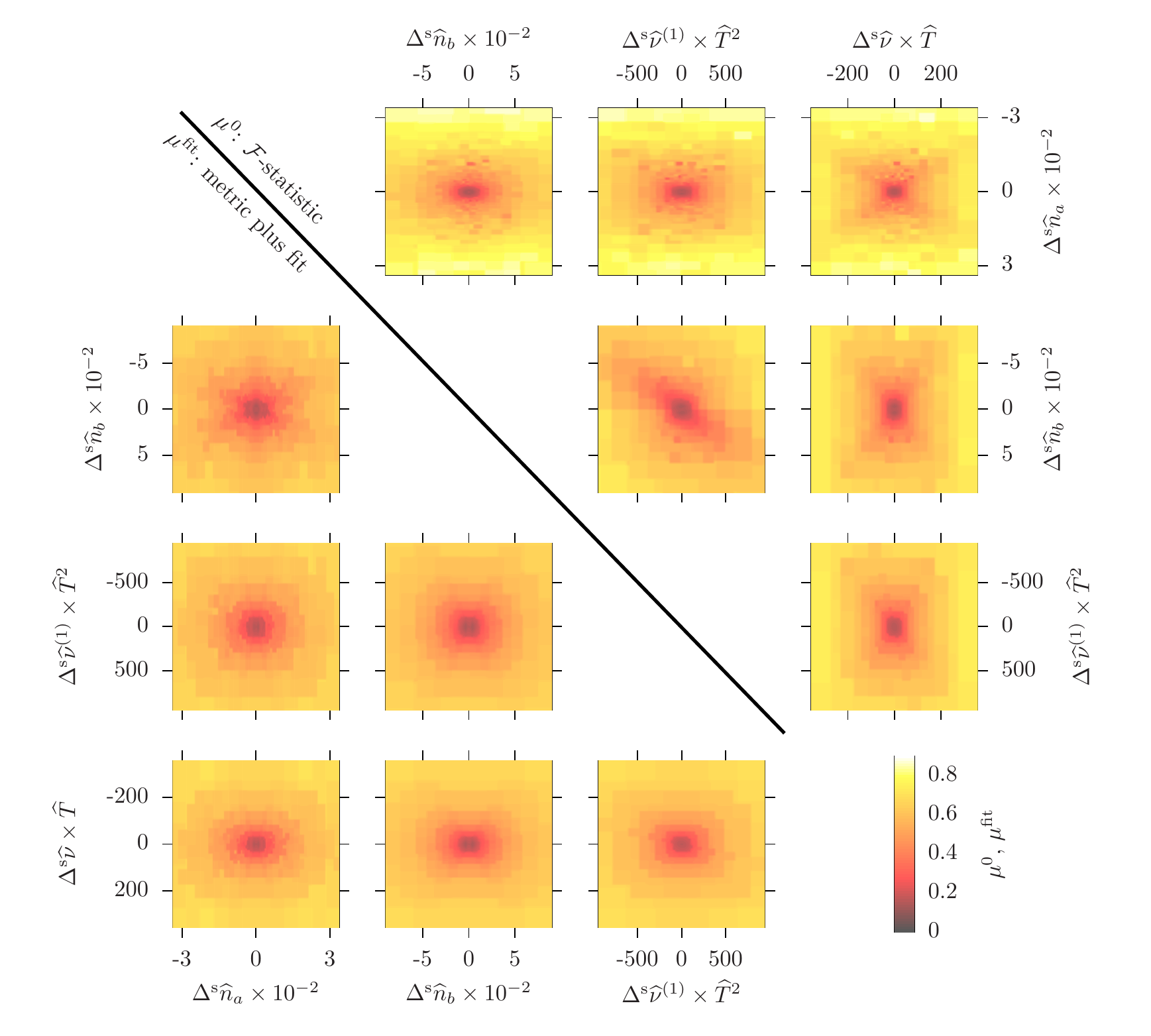}\label{fig:musptl_img_coh0p3_Tcoh1_Tsemi120}}\\
\subfloat[]{\includegraphics[width=\linewidth]{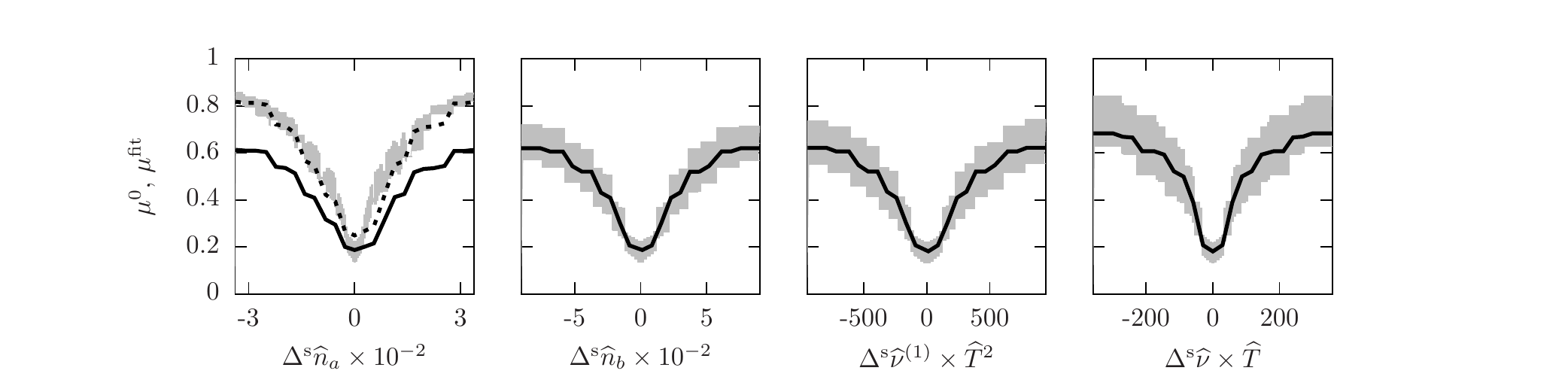}\label{fig:musptl_ax_coh0p3_Tcoh1_Tsemi120}}
\caption{\label{fig:musptl_coh0p1_Tcoh1_Tsemi240}
$\calF$-statistic mismatch as a function of offsets, between injected signals and their nearest semicoherent template, in the sky position $(\semi n\ua, \semi n\ub)$, spindown $\ndot[1]{\semi\nu}$ and frequency $\semi\nu$ parameters of the semicoherent metric, for $\coh T = 1$~day, $\semi T = 120$~days, and $\coh\mu\umax = 0.3$.
\protect\subref{fig:musptl_img_coh0p3_Tcoh1_Tsemi120}: $\calF$-statistic mismatch as a function of pairs of parameter offsets, all other offsets being $\sim 0$.
Above diagonal: $\calF$-statistic mismatch $\mu\Fstat$, given by Eq.~\eqref{eq:total-Fstat-mismatch}.
Below diagonal: mismatch $\mu\Ufit$ predicted by the metrics $(\semimat g, \cohsmat g)$ and improved by the empirical fit, given by Eq.~\eqref{eq:total-ssky-mismatch-fit}.
\protect\subref{fig:musptl_ax_coh0p3_Tcoh1_Tsemi120}: $\calF$-statistic mismatch as a function of individual parameter offsets, all other offsets being allowed to vary.
Gray shaded area: variation of $\calF$-statistic mismatch.
Black solid line: average over other offsets of the mismatch $\mu\Ufit$ predicted by the metrics $(\semimat g, \cohsmat g)$ and improved by the empirical fit.
Black dashed line: the same $\mu\Ufit$ multiplied by $0.8 / 0.6$; see the text for details.
}
\end{figure*}

\begin{figure*}
\subfloat[]{\includegraphics[width=\linewidth]{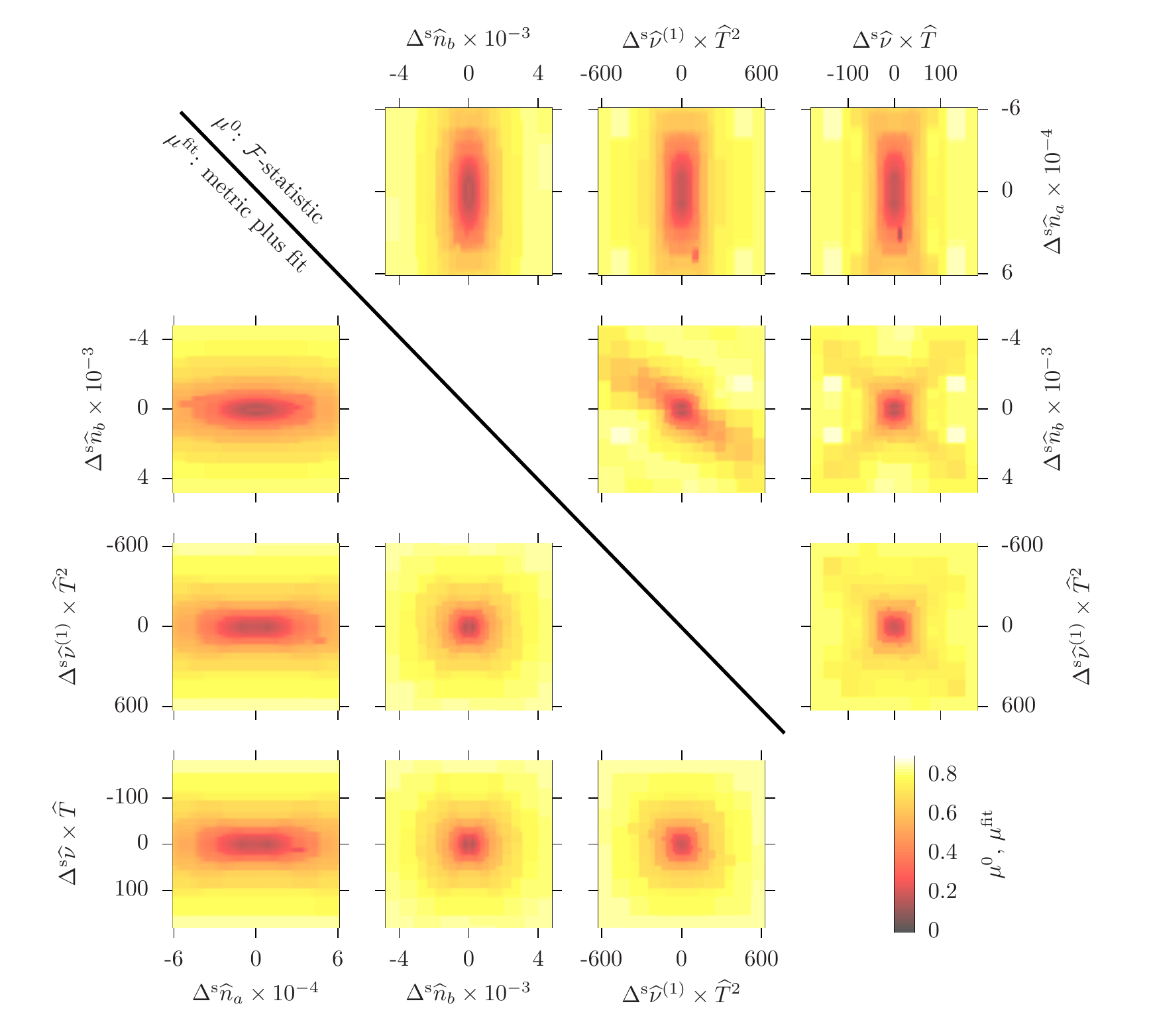}\label{fig:musptl_img_coh0p3_Tcoh3_Tsemi240}}\\
\subfloat[]{\includegraphics[width=\linewidth]{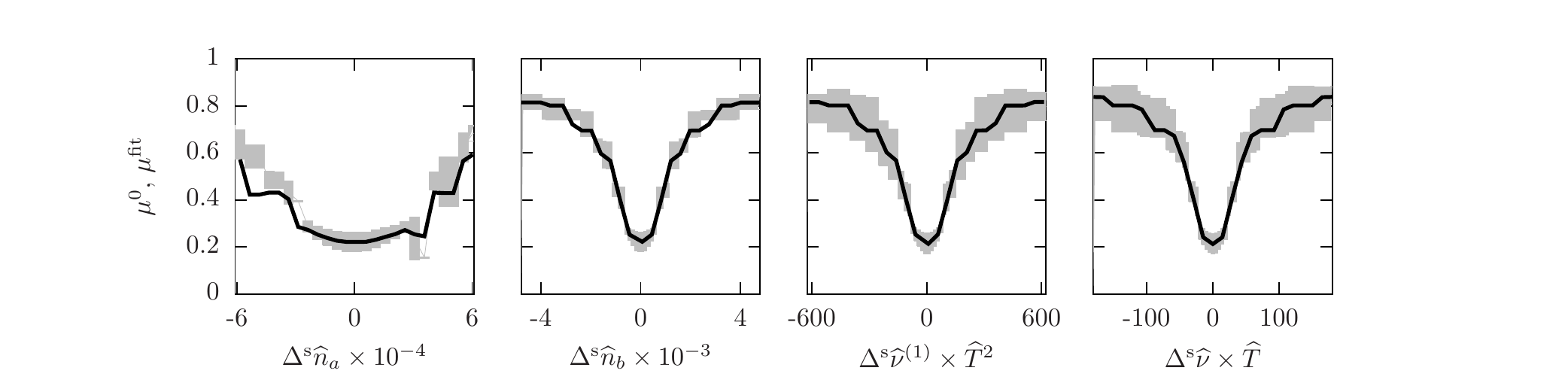}\label{fig:musptl_ax_coh0p3_Tcoh3_Tsemi240}}
\caption{\label{fig:musptl_coh0p3_Tcoh3_Tsemi240}
Same as Fig.~\protect\ref{fig:musptl_coh0p1_Tcoh1_Tsemi240}, but for $\coh T = 3$~days and $\semi T = 240$~days.
}
\end{figure*}

\begin{figure*}
\subfloat[]{\includegraphics[width=\linewidth]{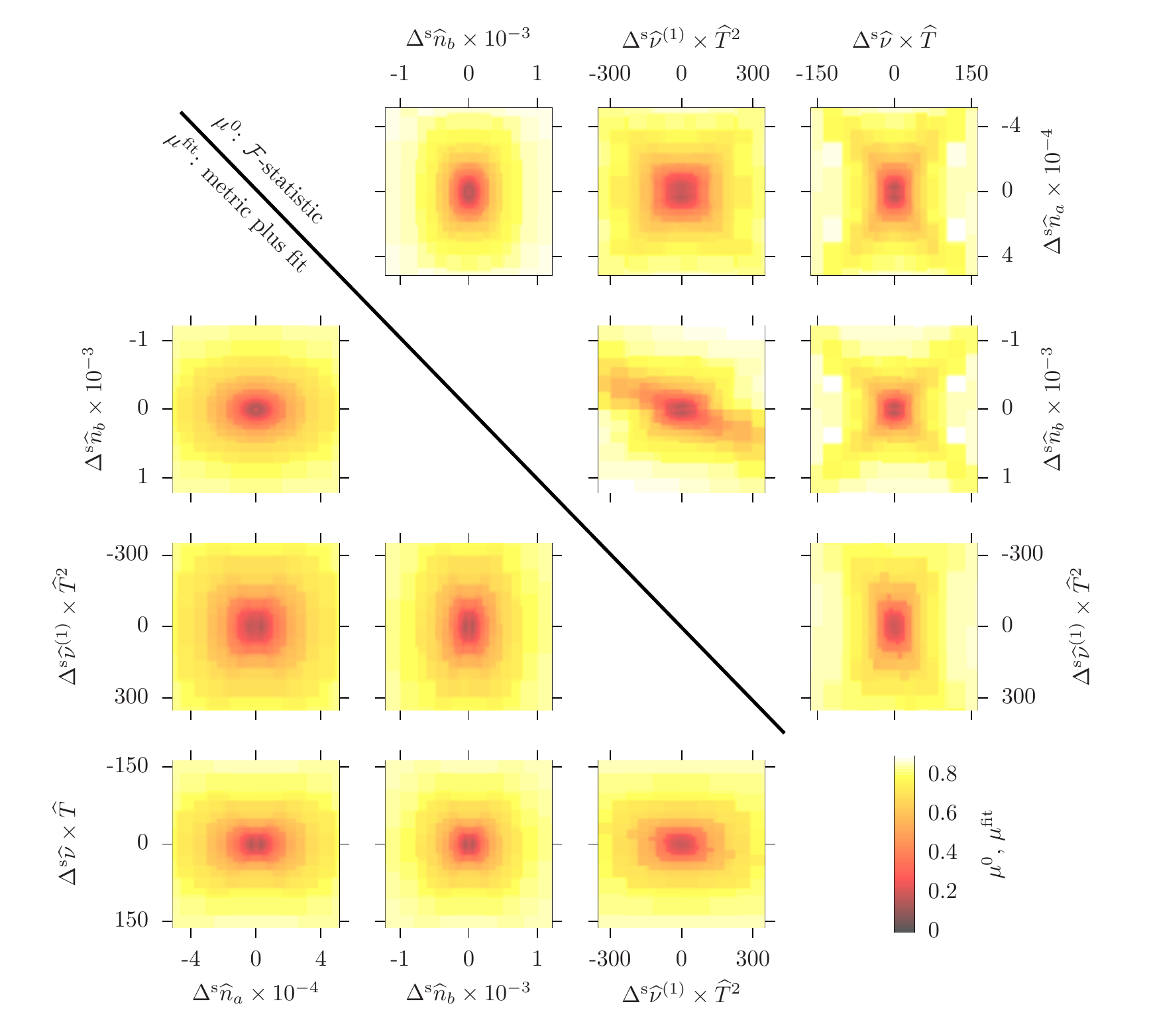}\label{fig:musptl_img_coh0p3_Tcoh5_Tsemi360}}\\
\subfloat[]{\includegraphics[width=\linewidth]{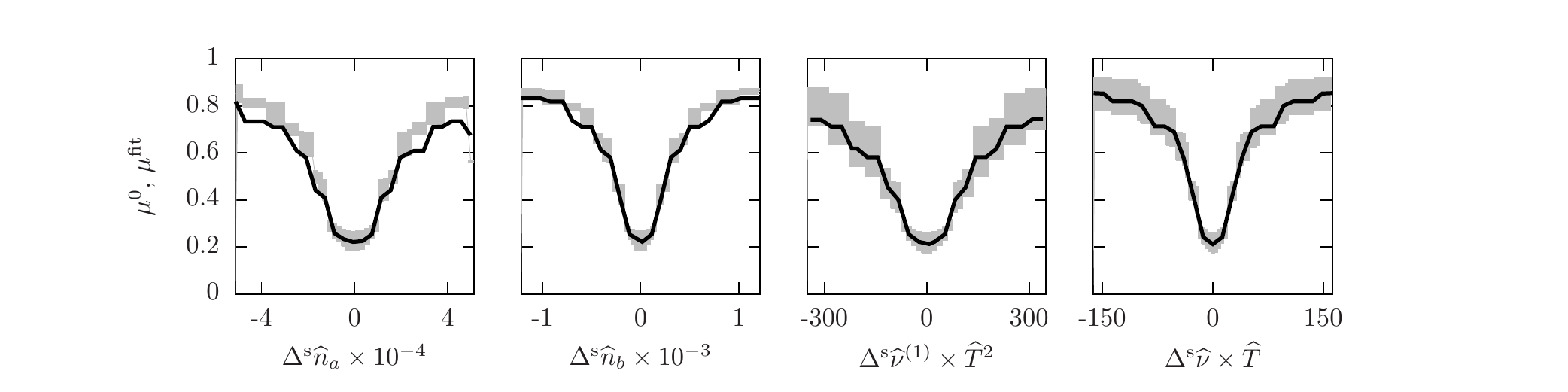}\label{fig:musptl_ax_coh0p3_Tcoh5_Tsemi360}}
\caption{\label{fig:musptl_coh0p3_Tcoh5_Tsemi240}
Same as Fig.~\protect\ref{fig:musptl_coh0p1_Tcoh1_Tsemi240}, but for $\coh T = 5$~days and $\semi T = 360$~days.
}
\end{figure*}

Equation~\eqref{eq:empr-mean-Fstat-fit}, derived in the previous section, gives us a tool $\mean{\mu\Fstat}\Ufit$ for predicting, with reasonable confidence, the mean $\calF$-statistic mismatch $\mean{\mu\Fstat}$, as a function of the mean metric mismatches $(\mean{\semi\mu}, \mean{\coh\mu})$, out to large $\mean{\mu\Fstat} \sim 0.9$.
One might also use this tool to improve the mismatch predicted of the metric between a signal $\sigvec\lambda$ and its nearest semicoherent and coherent templates $(\semivec\lambda, \cohsvec\lambda)$, by replacing $\mu$ [Eq.~\eqref{eq:total-ssky-mismatch}] with
\begin{multline}
\label{eq:total-ssky-mismatch-fit}
\mu\Ufit\Big( \coh T, \semi T, \sigvec\lambda, \semivec\lambda, \cohsvec\lambda \Big) \equiv \\ \mean{\mu\Fstat}\Ufit\Big( \coh T, \semi T, \semi\mu\big( \sig\diff\semivec\lambda \big), \mean{\coh\mu}\big( \semi\diff\cohsvec\lambda \big) \Big) \,,  
\end{multline}
where $\semi\mu$ is computed from $\sig\diff\semivec\lambda = \semivec\lambda - \sigvec\lambda$ via Eq.~\eqref{eq:semi-ssky-mismatch}, and $\mean{\coh\mu}$ is computed from the $\semi\diff\cohsvec\lambda = \cohsvec\lambda - \semivec\lambda$ via Eq.~\eqref{eq:avg-coh-ssky-mismatch}.
Here, the empirical fit provides the absolute \emph{scaling} of the $\calF$-statistic mismatch, while the metric provides \emph{directional} information, i.e.\ how the $\calF$-statistic mismatch changes in the direction of a vector $\sig\diff\semivec\lambda$ relative to some other vector $\sig\diff\semivec\lambda^{\prime}$.
Of course, given that Eq.~\eqref{eq:empr-mean-Fstat-fit} is fitted to the $\calF$-statistic mismatch averaged over signal and template parameters $(\sigvec\lambda, \semivec\lambda, \cohsvec\lambda)$, one would expect Eq.\eqref{eq:total-ssky-mismatch-fit} to not necessarily to be an accurate predictor of $\mu\Fstat$ for a \emph{particular} $(\sigvec\lambda, \semivec\lambda, \cohsvec\lambda)$.

In this section we examine the accuracy to which Eq.~\eqref{eq:total-ssky-mismatch-fit} models the $\calF$-statistic mismatch as a function of the search parameters of the semicoherent supersky metric: the sky position $(\semi n\ua, \semi n\ub)$, spindown $\ndot[1]{\semi\nu}$ and frequency $\semi\nu$.
The supersky metric coordinates are detailed in~\cite{Wette.Prix.2013a}; briefly, $(\semi n\ua, \semi n\ub)$ are components of the sky position vector $\semivec n$ in a preferred reference frame (which approaches the equatorial and ecliptic reference frames in the limit of short and long observation times respectively), and the $\ndot{\semi\nu}$ are equal to $\ndot{\semi f}$ plus a sky-position-dependent offset.
The numerical simulations described in Section~\ref{sec:calf-stat-mism} recorded the signal $\sigvec\lambda$ and template $(\semivec\lambda, \cohsvec\lambda)$ parameters for each computed $\mu\Fstat$; from these parameters the mismatch $\mu\Ufit$ predicted by the metric plus the empirical fit may be computed via Eq.~\eqref{eq:total-ssky-mismatch-fit}.

Figure~\ref{fig:musptl_coh0p1_Tcoh1_Tsemi240} compares $\mu\Fstat$ and $\mu\Ufit$, at fixed $\semi T = 120$~days, $\coh T = 1$~day, and $\coh\mu\umax = 0.3$, as a function of the parameter offsets between $\sigvec\lambda$ and $\semivec\lambda$, namely $\sig\diff\semi n\ua \equiv \semi n\ua - \sig n\ua$,
$\sig\diff\semi n\ub \equiv \semi n\ub - \sig n\ub$, $\sig\diff\ndot[1]{\semi\nu} \equiv \ndot[1]{\semi\nu} - \sig{[\ndot[1]\nu]}$, and $\sig\diff\semi\nu \equiv \semi\nu - \sig\nu$.
Figure~\ref{fig:musptl_img_coh0p3_Tcoh1_Tsemi120} plots mismatch as functions of distinct pairs of parameter offsets, all other parameter offsets being approximately zero~\footnote{ Since the numerical simulations of Section~\ref{sec:numer-simul-at} sample parameter offsets uniformly with respect to the metric, it is unlikely that the offset in any parameter will be precisely zero. For the purpose of these plots, offsets which are $\lesssim 1$\% of the total range of offsets generated by the simulations are treated as zero. }.
For example, the top-right subplot shows $\mu\Fstat$ as a function of $\sig\diff \semi n\ua$ and $\sig\diff \semi\nu$ with $\sig\diff \semi n\ub \sim \sig\diff \ndot[1]{\semi\nu} \sim 0$.
Note that the plots of $\mu\Fstat$ (above the diagonal) are transposed images of the corresponding plots of $\mu\Ufit$ (below the diagonal).

As expected, mismatches are zero (darkest color) when signal and template are perfectly matched, and increase monotonically (to lighter colors) in response to any offsets between signal and template parameters.
The behavior of the $\calF$-statistic mismatch $\mu\Fstat$ as a function of offsets is generally well-modeled by $\mu\Ufit$, as shown by the similarity of the corresponding subplots in Fig.~\ref{fig:musptl_img_coh0p3_Tcoh1_Tsemi120}.
This indicates that $\mu\Ufit$ is a reasonable model for $\mu\Fstat$ even out to large $\mu\Fstat \sim 0.9$.
It also implies that, while the derivation of the parameter-space metric (see Section~\ref{sec:background}) loses the correct absolute scaling of the $\calF$-statistic mismatch at large $\mu\Fstat$, it does retain the correct directional information.

The one exception to the above, in Fig.~\ref{fig:musptl_img_coh0p3_Tcoh1_Tsemi120}, is the mismatch behavior with respect to offsets in the sky position parameter $\semi n\ua$.
Comparing the first row of subplots (above the diagonal) in Fig.~\ref{fig:musptl_img_coh0p3_Tcoh1_Tsemi120} with the first column (below the diagonal), we see that $\mu\Fstat$ increases more quickly as a function of $\sig\diff\semi n\ua$ than does $\mu\Ufit$.
This effect is more readily apparent in Fig.~\ref{fig:musptl_ax_coh0p3_Tcoh1_Tsemi120}, where we plot $\mu\Fstat$ and $\mu\Ufit$ as functions of individual parameter offsets -- namely $\sig\diff\semi n\ua$, $\sig\diff\semi n\ub$, $\sig\diff\ndot[1]{\semi\nu}$, and $\sig\diff\semi\nu$ -- and where the other 3 offsets are allowed to vary over their simulated ranges.
We see that, as a function of $\sig\diff\semi n\ua$ (left-most subplot in Fig.~\ref{fig:musptl_ax_coh0p3_Tcoh1_Tsemi120}), $\mu\Fstat$ (gray shaded area) increases more rapidly that $\mu\Ufit$ (black solid line); at $\sig\diff\semi n\ua \sim \pm 3{\times}10^{-2}$, $\mu\Fstat \sim 0.8$ whereas $\mu\Ufit \sim 0.6$.

The reason for this discrepancy is likely due to numerical issues in computing the supersky metric, which requires the eigenvalues of the sky--sky block of a precursor metric~\cite{Wette.Prix.2013a}.
At $\coh T \sim 1$~day, however, the precursor metric is highly ill-conditioned~\cite{Prix.2007a,Wette.Prix.2013a}, which may lead to inaccurate computation of the eigenvalues.
It is likely that, in this instance, the $\semi n\ua$--$\semi n\ua$ component of the supersky metric, which is proportional to the largest eigenvalue, has been inaccurately computed.
To illustrate this, we re-plot $\mu\Ufit$ re-scaled by $0.8 / 0.6$, and see that the re-scaled $\mu\Ufit$ now follows $\mu\Fstat$ closely.
This indicates a systematic error in the supersky metric as a function of $\semi n\ua$, which we surmise is most likely due to inaccurate computation of the largest eigenvalue.

Figures~\ref{fig:musptl_coh0p3_Tcoh3_Tsemi240} and~\ref{fig:musptl_coh0p3_Tcoh5_Tsemi240} compare $\mu\Fstat$ and $\mu\Ufit$ in a similar manner to Fig.~\ref{fig:musptl_coh0p1_Tcoh1_Tsemi240}, but at larger $\coh T$ and $\semi T$.
As $\coh T$ increases to 3~days (Fig.~\ref{fig:musptl_coh0p3_Tcoh3_Tsemi240}) and 5~days (Fig.~\ref{fig:musptl_coh0p3_Tcoh5_Tsemi240}), the discrepancy between $\mu\Fstat$ and $\mu\Ufit$ as a function of $\sig\diff\semi n\ua$ largely disappears; compare the left-most plots in Figs.~\ref{fig:musptl_ax_coh0p3_Tcoh1_Tsemi120}, \ref{fig:musptl_ax_coh0p3_Tcoh3_Tsemi240}, and~\ref{fig:musptl_ax_coh0p3_Tcoh5_Tsemi360}.
This is expected, since the precursor metric becomes better-conditioned as $\coh T$ increases, and therefore the computation of the eigenvalues becomes more reliable.

Some discrepancies between $\mu\Fstat$ and $\mu\Ufit$ as a function of $\sig\diff\semi n\ub$ and $(\sig\diff\ndot[1]{\semi\nu}, \sig\diff\semi\nu)$ are evident in Figs.~\ref{fig:musptl_img_coh0p3_Tcoh3_Tsemi240} and~\ref{fig:musptl_img_coh0p3_Tcoh5_Tsemi360}.
Compare the second row of subplots (above the diagonal), which display banded and/or cross-shaped features, with the second column of subplots (below the diagonal), where those features are absent.
This is likely due to assumptions made in deriving the supersky metric~\cite{Wette.Prix.2013a}.
Briefly, the metric tries to model the orbital motion of the Earth by a second-order Taylor expansion, which can then be absorbed into the frequency and spindown parameters; a small component of the residual orbital motion, i.e.\ that component which cannot be modeled by a Taylor expansion, is then discarded.
This introduces an error into the supersky metric which is generally small, but is also proportional to parameter offsets; hence at large $\mu\Fstat$, and hence $\sig\diff\semi n\ub$ and $(\sig\diff\ndot[1]{\semi\nu}, \sig\diff\semi\nu)$, the effect of this error is magnified.
Nevertheless, as can be seen from Figs.~\ref{fig:musptl_ax_coh0p3_Tcoh3_Tsemi240} and~\ref{fig:musptl_ax_coh0p3_Tcoh5_Tsemi360}, $\mu\Ufit$ generally tracks the average $\mu\Fstat$ as a function of individual offsets.

Finally, note the small feature $\mu\Fstat$ in the left-most subplot in Fig.~\ref{fig:musptl_ax_coh0p3_Tcoh3_Tsemi240} at $\sig\diff\semi n\ua \sim +3{\times}10^{-4}$, which does not appear at $\sig\diff\semi n\ua \sim -3{\times}10^{-4}$; one would expect $\mu\Fstat$ to be insensitive to the sign of $\sig\diff\semi n\ua$.
This is likely due to a minor issue in the implementation of the transformation from supersky $(\semi n\ua, \semi n\ub)$ to physical $(\alpha, \delta)$ coordinates, where for very small template banks [i.e.\ small $(\coh T, \semi T)$ and high $(\coh\mu\umax, \semi\mu\umax)$] neighboring templates in $(\semi n\ua, \semi n\ub)$ can end up in opposite hemispheres when mapped to $(\alpha, \delta)$.
This is not an issue for template banks of realistic densities.

\section{Discussion}\label{sec:discussion}

The study described in this paper was motivated by the realization that, under realistic computing cost constraints, a semicoherent search based on the parameter-space metric of~\cite{Wette.2015a} could not be performed with maximum semicoherent mismatches within the range of validity of the metric, i.e.\ $\lesssim 0.4$.
This situation was not realized by previous work of the semicoherent metric~\cite{Brady.Creighton.2000a,Pletsch.2010a} since those works do not accurately predict the number of semicoherent templates once $\semi T \gtrsim 20$~days~\cite{Wette.2015a}.
Other semicoherent search methods~\cite[e.g.][]{Krishnan.etal.2004a} do not use an explicit metric to describe the parameter space.

The key finding of this paper is that the mean $\calF$-statistic mismatch increases only slowly with $(\coh\mu\umax, \semi\mu\umax)$; see Fig.~\ref{fig:mutwoF_fit}.
As discussed in Section~\ref{sec:introduction}, it appears likely that an all-sky semicoherent search based on the metrics of~\cite{Wette.Prix.2013a,Wette.2015a} would have to operate at a high $\semi\mu\umax$, in order to satisfy reasonable computing cost constraints.
The results presented here give us some confidence that, despite a high $\semi\mu\umax$, the mean $\calF$-statistic mismatch of such a search, and hence its sensitivity, would remain competitive.
A thorough examination of the sensitivity of such a search is planned for future work.

Note, however, that the relationship between $\mean{\mu\Fstat}$ and $(\coh\mu\umax, \semi\mu\umax)$ need not be known \emph{a priori} in order to implement a semicoherent search based on~\cite{Wette.Prix.2013a,Wette.2014a,Wette.2015a}.
First, a semicoherent template bank can be constructed for any $\semi\mu\umax$, however large; it is only when estimating the sensitivity of such a search that one must correctly translate between $\semi\mu\umax$ and $\mean{\mu\Fstat}$.
Second, the nearest coherent template in each segment, used to compute the summed $\calF$-statistic of Eq.~\eqref{eq:semi-Fstat-def}, are determined by the coherent metrics in each segment; preliminary search optimization has found that, while $\semi\mu\umax$ may be large, $\coh\mu\umax$ is likely to remain small, i.e.\ within the range of validity of the coherent metric.

The search setup optimization procedure of~\cite{Prix.Shaltev.2012a} assumes that the relationship between $\mean{\mu\Fstat}$ and $(\coh\mu\umax, \semi\mu\umax)$ is strictly proportional; see Eq.~(23) of that paper.
A possible extension to the method of~\cite{Prix.Shaltev.2012a} could be to allow instead an arbitrary functional relationship, e.g.\ that given by Eq.~\eqref{eq:empr-mean-Fstat-fit}.
This could potentially lead to improved optimal search setups, since the optimization would be aware that increasing $(\coh\mu\umax, \semi\mu\umax)$ do not increase $\mean{\mu\Fstat}$, and hence degrade sensitivity, as much as previously assumed.

When deriving the parameter-space metric, it is standard to assume~\cite[e.g.][]{Prix.2007a,Wette.Prix.2013a} an ideal case where the noncentrality parameter $\rho^2$ (which is proportional to the signal-to-noise-ratio~\cite{Wette.2015a}) is large, i.e. the data being searched contains little or no noise.
The addition of noise will certainly reduce the recovered signal-to-noise-ratio (and consequentially the noncentrality parameter), whether at a mismatched template [$\rho^2(\vec\calA, \sigvec\lambda; \cohsvec\lambda)$] or at perfect match [$\rho^2(\vec\calA, \sigvec\lambda; \sigvec\lambda)$].
The mismatch is, however, a \emph{ratio} of noncentrality parameters [Eq.~\eqref{eq:coh-Fstat-mismatch-def}], so a constant reduction in noncentrality parameter does not affect the mismatch.
It is probable that noise would affect $\rho^2(\vec\calA, \sigvec\lambda; \cohsvec\lambda)$ and $\rho^2(\vec\calA, \sigvec\lambda; \sigvec\lambda)$, since slightly different data would be used in the computation of each.
We expect that such an effect would average out over a large number of templates, however, and therefore hypothesize that the parameter-space metric would still provide a useful prediction of the \emph{mean} measured mismatch, even in the realistic case of noisy data.

\acknowledgments

I thank Reinhard Prix and Maria Alessandra Papa for valuable discussions.
Numerical simulations were performed on the ATLAS computer cluster of the Max-Planck-Institut f\"ur Gravitationsphysik.
This paper has document number LIGO-P1600162.

\appendix

\section{Standard deviation of $\calF$-statistic mismatch}\label{sec:stand-devi-calf}

\begin{table}
\begin{tabular*}{\linewidth}{l@{\extracolsep{\fill}}r@{\extracolsep{0pt}}l@{\extracolsep{\fill}}r@{\extracolsep{0pt}}l@{\extracolsep{\fill}}r@{\extracolsep{0pt}}l@{\extracolsep{2\tabcolsep}}}
\hline\hline
 & \multicolumn{6}{c}{Coefficients} \\
$n$ & \multicolumn{2}{c}{$1$} & \multicolumn{2}{c}{$2$} & \multicolumn{2}{c}{$3$} \\
\hline
$A\Ufit_{n}$ & $0$ & $.88568$ & $2$ & $.4022$ & $2$ & $.0377$ \\
$B\Ufit_{n}$ & $-0$ & $.35794$ & $-0$ & $.38261$ & $0$ & $.091035$ \\
$C\Ufit_{n}$ & $-0$ & $.11033$ & $-0$ & $.01705$ & $0$ & $.059349$ \\
$D\Ufit_{n}$ & $2$ & $.0924$ & $2$ & $.243$ & $2$ & $.1754$ \\
$E\Ufit_{n}$ & $-0$ & $.0024717$ & $0$ & $.29952$ & $0$ & $.019583$ \\
$F\Ufit_{n}$ & $0$ & $.01496$ & $0$ & $.091887$ & $0$ & $.90241$ \\
$G\Ufit_{n}$ & $-1$ & $.1985$ & $-2$ & $.1845$ & $-7$ & $.3365$ \\
$H\Ufit_{n}$ & $-0$ & $.094061$ & $-2$ & $.6435$ & \multicolumn{2}{c}{$0$} \\
\hline\hline
\end{tabular*}
\caption{\label{tab:stdv_fit_coefficients}
Coefficients of fit for Equation~\eqref{eq:empr-stdv-Fstat-fit}.
}
\end{table}

We also find an empirical fit to the standard deviations $\stdv{\mu\Fstat}$ of the $\calF$-statistic mismatch.
The fit is a function of $\coh T$, $\semi T$, and the standard deviations $\stdv{\coh\mu}$ and $\stdv{\semi\mu}$ of the coherent and semicoherent metric mismatches; ratios of these quantities to the maxima $\semi\mu\umax$ and $\coh\mu\umax$ respectively are listed in Table~\ref{tab:lattice_mean_stdv_table}.
The empirical fit is given by
\begin{equation}
\label{eq:empr-stdv-Fstat-fit}
\begin{split}
\stdv{\mu\Fstat}\Ufit &= N\Ufit \sum_{n=1}^{3} \frac{1}{n}
\exp\left[ A\Ufit_{n} + B\Ufit_{n} \, \frac{\semi T}{\text{year}} + C\Ufit_{n} \, \frac{\coh T}{\text{day}} \right] \\
&\quad\times \exp\Big[ -\big( D\Ufit_{n} + E\Ufit_{n} \stdv{\semi\mu} + F\Ufit_{n} \stdv{\coh\mu} \big)^2 \Big] \\
&\quad\times \Big[ 1 - \exp\big( G\Ufit_{n} \stdv{\semi\mu} + H\Ufit_{n} \stdv{\coh\mu} \big) \Big] \,,
\end{split}
\end{equation}
where the fitted coefficients $N\Ufit = 2.9101$, and $A\Ufit_{n}$ through $H\Ufit_{n}$ are listed in Table~\ref{tab:stdv_fit_coefficients}.
Over the 576 values of $\stdv{\mu\Fstat}$ used for fitting, the root-mean-square relative error to $\stdv{\mu\Fstat}\Ufit$ was minimized to $\lesssim 9$\%.

\bibliography{paper}

\end{document}